\documentclass{article}
\usepackage[english]{babel}
\usepackage{mathtools}
\usepackage{hyperref}

\mathtoolsset{showonlyrefs=true}
\addto\extrasenglish{    }

\usepackage[margin=1.3in]{geometry}

\usepackage{graphicx}
\usepackage{bmpsize}

\usepackage{enumitem}
\usepackage{url}

\usepackage{amsfonts}
\usepackage{amsmath}
\usepackage{hyperref}
\usepackage{xargs}                      
\usepackage[pdftex,dvipsnames]{xcolor}  
\usepackage[colorinlistoftodos,prependcaption,textsize=tiny]{todonotes}
\usepackage[normalem]{ulem}
\usepackage{longtable}
\newcommandx{\fatemeh}[2]{{\uline{#1}}\textcolor{pink}{(#2)}}
\newcommandx{\handan}[2]{{\uline{#1}}\textcolor{red}{(#2)}}
\newcommandx{\eray}[2]{{\uline{#1}}\textcolor{blue}{(#2)}}
\newcommandx{\jeff}[2]{{\uline{#1}}\textcolor{green}{(#2)}}
\newcommandx{\al}[2]{{\uline{#1}}\textcolor{cyan}{(#2)}}
\newcommandx{\sergey}[2]{{\uline{#1}}\textcolor{gray}{(#2)}}
\newcommandx{\alfonso}[2]{{\uline{#1}}\textcolor{purple}{(#2)}}
\newcommandx{\syed}[2]{{\uline{#1}}\textcolor{orange}{(#2)}}
\newcommandx{\ximin}[2]{{\uline{#1}}\textcolor{red}{(#2)}}
\newcommandx{\leon}[2]{{\uline{#1}}\textcolor{brown}{(#2)}}

\newcommand{\Nom}{\ensuremath{\mathcal{N}}} 
\newcommand{\Val}{\ensuremath{\mathcal{V}}} 
\newcommand{\nval}{\ensuremath{n_{val}}} 
\newcommand{\Can}{\ensuremath{\mathcal{C}}} 
\newcommand{\nom}{\ensuremath{n}} 
\newcommand{\val}{\ensuremath{v}} 
\newcommand{\col}{\ensuremath{c}} 
\newcommand{\Par}{\ensuremath{P}} 
\newcommand{\Col}{\ensuremath{\mathcal{C}}} 
\newcommand{\slot}{\ensuremath{sl}} 
\newcommand{\ep}{\ensuremath{e}} 
\newcommand{\D}{\ensuremath{\Delta}}
\newcommand{\sk}{\ensuremath{\mathsf{sk}}} 

\title{Overview of Polkadot and its Design Considerations}

\author{
  Jeff Burdges$^1$, Alfonso Cevallos$^1$, Peter Czaban$^1$\\
  \and
  Rob Habermeier$^2$, Syed Hosseini$^1$, Fabio Lama$^1$,\\
  \and
  Handan K{\i}l{\i}n\c{c} Alper$^1$, Ximin Luo$^1$, Fatemeh Shirazi$^1$, \\
  \and
  Alistair Stewart$^1$, Gavin Wood$^{1,2}$\\
  \\
  \texttt{$^1$ Web3 Foundation},\\
  \texttt{$^2$ Parity Technologies}
}

\begin{document}

\maketitle

\begin{abstract}
In this paper we describe the design components of the heterogenous multi-chain protocol Polkadot and explain how these components help Polkadot address some of the existing shortcomings of blockchain technologies.
At present, a vast number of blockchain projects have been introduced and employed with various features that are not necessarily designed to work  with each other.  This makes it difficult for users to utilise a large number of applications on different blockchain projects. Moreover, with the increase in number of projects the security that each one is providing individually becomes weaker.
Polkadot aims to provide a scalable and interoperable framework for multiple chains with pooled security that is achieved by the collection of components described in this paper.
\end{abstract}

\tableofcontents
\newpage
\section{Introduction}\label{sec:intro}
The Internet was originally designed for and built upon decentralised protocols such as TCP/IP, however, its commercialisation has led to the centralisation of all popular web applications today. We refer not to any centralisation of physical infrastructure, but rather to the logical centralisation of power and control over the infrastructure. Two prominent examples are giant corporations like Google and Facebook: while they maintain servers all around the world in a physically decentralised fashion, these are ultimately controlled by a single entity.

A central entity controlling a system poses many risks for everyone. For example, they can stop the service at any moment, can sell users' data to third parties, and manipulate how the service is working without the users' agreement. This is in particular critical to users who heavily rely upon these services for business or private purposes.

With all the current interest related to personal data ownership, there is a growing need for a better security, freedom and control for net users, and with this a countermovement for more decentralised applications where no single entity controls the system. This tendency towards decentralisation is not new; it has been used in a number of areas of the development of web and other systems, such as the free software movement. 

Blockchains are a recent technology proposed to address these problems, in the quest to build a decentralised web. However, it can only compete with the centralised web if it is usable for masses of end-users. One important aspect of this is that separate applications must be able to interact, otherwise each will become isolated and not adopted by as many users. Having to build such an interoperability mechanism introduces new challenges, many of which are missing in the centralised model because of the fundamental differences in the trust model between the two paradigms.
For example, Bitcoin\cite{nakamoto2008bitcoin} and Ethereum\cite{buterin2014ethereum} are proof-of-work (PoW) blockchains where security relies on assumptions about processing power; alternative proof-of-stake (PoS) systems' security instead rely on incentives and the ability to destroy security deposits. These differences present difficulties for one blockchain to trust another. Another challenge that blockchain technologies need to tackle is scalability. Existing blockchain systems generally have high latency and can only have tens of transactions per second \cite{scaling}, whereas credit card companies such as Mastercard or Visa carry out thousands of transactions per second \cite{Visa2020}.

One prominent solution to scalability for blockchains is to run many chains in parallel, often called sharding. 
Polkadot is a multi-chain system that aims to gather the security power of all these chains together in a shared security system. It was first introduced in 2016 by Gavin Wood \cite{2016:Wood:Polkadot}, and in this paper we expand on the details.

Briefly: Polkadot utilises a central chain called the \emph{relay chain} which communicates with multiple heterogeneous and independent sharded chains called \emph{parachains} (portmanteau of “parallel chains”). The relay chain is responsible for providing shared security for all parachains, as well as trust-free interchain transactability between parachains. In other words, the issues that Polkadot aims to address are those discussed above: interoperability, scalablility, and weaker security due to splitting the security power.

\paragraph{Paper Organisation} In the next section, we give a synopsis of the Polkadot network including its external interface with client parachains that we expand on in subsequent sections. We review preliminary information such as description of roles of Polkadot's participants and our adversary model in Section \ref{sec:preliminaries}. We explain what subprotocols and components of Polkadot try to achieve in Section \ref{sec:components} and then continue to review them in detail, including low-level cryptographic and networking primitives. Finally we shortly discuss some future work in Section \ref{sec:futurework}. In the appendices, we review relevant work such as a comparison to other multi-chain system  \ref{sec:comparison}, a short description of an interoperability scheme to bridge to external chains \ref{sec:bridge}, a secure execution schemes \ref{sec:SPREE} that we will use for  messaging, and a glossary with Polkadot-specific terms in Table \ref{t:time}. 




\section{Synopsis}\label{sec:summary}
The aim of this section is to describe the main functionality of Polkadot without going into details about the design considerations and reasoning.

The Polkadot system consists of a single open collaborative decentralised network called the relay chain, that interacts with many other external chains run in parallel called \emph{parachains}. From a high-level perspective, parachains are clients of the relay chain, which provides a security service to these clients, including secure communication. That is the sole purpose of the relay chain; parachains are the entities providing application-level functionality, such as cryptocurrencies.

The internal details of parachains are not a concern of the relay chain; parachains need only adhere to the interface we specify. Some of these expectations are natural components of blockchains, hence the naming. However, other non-blockchain systems may also run as a Polkadot parachain as long as they satisfy the interface. This is described below: relevant parts are \uline{underlined}.

These aspects may be abbreviated as, \emph{Polkadot is a scalable heterogeneous multi-chain}.

\subsection{Security model}

We assume that parachains are running as external untrusted clients of the relay chain and that the relay chain only deals with parachains via an interface and does not make assumptions about their internals. For example, internally they may be permissioned or open; if some internal users subvert the parachain, from Polkadot's viewpoint the entire parachain (as a single client entity) is malicious.

The Polkadot relay chain is designed to deal with a level of malicious behaviour internally, as a requirement of being an open decentralised network. Specific individual nodes are untrusted, but an indeterminable subset of nodes lower-bounded in size are trusted, and the protocol works to ensure that the relay chain externally as a whole is trustable. See section \ref{sec:security_model} for details.

\subsection{Nodes and roles}

The Polkadot relay chain network consists of nodes and roles. Nodes are the network-level entities physically executing the Polkadot software, and roles (Section \ref{sec:roles}) are protocol-level entities performing a particular purpose. Nodes may perform multiple roles.

On the network level, the relay chain is open. Any node may run the software and participate as any of these types of nodes:

\begin{enumerate}
	\item Light client - retrieves certain user-relevant data from the network. The availability of light clients is irrelevant - they don't perform a service for others.
	\item Full node - retrieves all types of data, stores it long-term, and propagates it to others. Must be highly available.
	\begin{enumerate}
		\item \hyperref[sec:net_sentry]{Sentry node} - publicly-reachable full nodes that perform trusted proxying services for a private full node, run by the same operator.
	\end{enumerate}
	Sometimes we refer to a \emph{full node} of a parachain. In the abstract sense for non-blockchain parachains, this means that they participate in it to a sufficient degree that they can verify all data passing through it.
\end{enumerate}

Beyond distributing data, relay-chain nodes may perform certain protocol-level roles listed next. Some of these roles have restrictions and conditions associated with them:

\begin{enumerate}
	\item Validator - performs the bulk of the security work. Must be a full node of the relay chain. Interacts with parachain collators, but need not participate in a parachain as a full node. 
	\item Nominator - stakeholder who backs and selects validator candidates (Section \ref{sec:validators}). This can be done from a light client, and they need not have any awareness of parachains.
\end{enumerate}

Parachains may decide their own internal network structure, \uline{but are expected to interact with Polkadot via the following roles}:

\begin{enumerate}
	\item Collator\footnotemark[1] - collects and submits parachain data to the relay chain, subject to protocol rules described below. They are chosen as defined by the parachain, and must be full nodes of it.
	\item Fishermen - performs additional security checks on the correct operation of the parachain, on behalf of the relay chain who supplies a reward. This role is self-assigned and reward-incentivized, and must be a full node of the parachain.
\end{enumerate}

\footnotetext[1]{Validators are in some sense a collator of the relay chain network, in that they collect extrinsics (e.g. transactions) within the relay chain network. However we typically only refer to them as validators even when performing these tasks, and the term \emph{collator} is reserved for parachain collators only.}

\subsection{Protocol}

The Polkadot relay chain protocol, including interaction with parachains, works as follows.

\begin{enumerate}
	\item For each parachain:
	
	\begin{enumerate}
		\item \uline{Collators watch the progress of the block-producing and consensus protocols}, steps (2) and (5) respectively below, e.g. by participating in the relay chain as a full node. Based on what they think is the latest relay chain block that will most likely be finalised, they build on top of the latest parachain block (or other data) that would be finalised by it.
		\item \uline{Collators sign data building on top of said latest parachain block, and submit it possibly indirectly}, to the validators assigned to their parachain (\emph{parachain validators} for short), for inclusion in the relay chain. Ideally they submit a unique one, to help performance.
		\item The parachain validators decide which parachain block to support, and presents relevant data of it as a parachain's next \emph{candidate} to be added to the next relay chain block.
	\end{enumerate}
	
	\item A block-producing validator collects candidates from all parachains, and puts this collection along with any recent relay chain extrinsics into a relay chain head block. (Section \ref{sec:babe}). For performance, this does not contain the full data from all parachains, but only metadata and partial data, including security-related metadata.
	
	In the unfavourable case, this can result in forks, resolved later in step (5). This subprotocol is designed so that even with forks, participants have an idea of the block most likely to be finalised, similar to Proof-of-Work protocols.
	
	\item A subprotocol is run to ensure that the full data is indeed available, including and distributing it to various other relay-chain nodes. (Section \ref{sec:validity-and-availability}).
	
	\item Data submitted from a parachain might include indications that they are sending messages to another parachain, including metadata to facilitate this. This is now included on the relay chain head(s), so recipient parachains are aware of which new messages have been sent to them. \uline{They now retrieve the message bodies from the sending parachains}. (Section \ref{sec:XCMP}).
	
	\item Validators submit their votes on the block and finalises it, resolving any forks to a single head. (Section \ref{sec:grandpa}). These votes are added to the relay chain blocks.
	
\end{enumerate}


The rest of the paper expands on the above - roles next in Section \ref{sec:preliminaries}, and protocol subcomponents in Section \ref{sec:components}.

\section{Preliminaries}\label{sec:preliminaries}


In this section we describe in more detail the different entities involving in the running of Polkadot, as well as our security model of these entities. This forms the design context for understanding the protocol design described later, including how it works and why it is designed that way.




\subsection{Roles}\label{sec:roles}
Nodes which run Polkadot network assume different roles and functions that we introduce next.


\paragraph{Validators:}\label{par:validators} A validator is the highest in charge and helps seal new blocks on the Polkadot network. The validator’s role is contingent upon a sufficiently high bond being deposited, though we allow other bonded parties to nominate one or more validators to act for them and as such some portion of the validator’s bond may not necessarily be owned by the validator itself but rather by these nominators.
A validator must run a relay-chain client implementation with high availability and bandwidth. At each block the node must be ready to accept the role of ratifying a new block on some parachain, and may be required to double check a few more. This process involves receiving, validating and republishing candidate blocks. The parachain assignment is random and changes frequently. Since the validator cannot reasonably be expected to maintain a fully-synchronised database of all parachains, the task of devising a suggested new parachain block will be delegated to a third-party, known as a collator.
Once all new parachain blocks have been properly ratified by their appointed validator subgroups, validators must then ratify the relay-chain block itself. This involves updating the state of the transaction queues (essentially moving data from a parachain’s output queue to another parachain’s input queue), processing the transactions of the ratified relay-chain transaction set and ratifying the final block, including the final parachain changes.
A validator provably not fulfilling their role will be slashed i.e. part or all of their bond will be taken.
In some sense, validators are similar to the mining pools of current PoW blockchains.

\paragraph{Nominators:}\label{par:nominators} A nominator is a stake-holding party who contributes to the security bond of a validator. They have no additional role except to place risk capital and as such to signal that they trust a particular validator (or set thereof) to act responsibly in their maintenance of the network. They receive a pro-rata increase or reduction in their deposit according to the bond’s growth to which they contribute. Together with collators, next, nominators are in some sense similar to the miners of the present-day PoW networks.

\paragraph{Collators: }\label{par:collators} Transaction collators (collators for short) are parties who assist validators in producing valid parachain blocks. They maintain a “full-node” for a particular parachain; meaning that they retain all necessary information to be able to author new blocks and execute transactions in much the same way as block producers do on current blockchains. Under normal circumstances, they will collate and execute transactions to create an unsealed block, and provide it, together with a proof of validity, to one or more validators presently responsible for proposing a parachain block.


\paragraph{Fishermen:} \label{par:fishermen} Unlike the other two active parties, fishermen are not directly related to the block-authoring process. Rather they are independent “bounty hunters” motivated by a large one-off reward. Precisely due to the existence of fishermen, we expect events of misbehaviour to seldom happen, and when they do only due to the bonded party being careless with secret key security, rather than through malicious intent. The name comes from the expected frequency of reward, the minimal requirements to take part and the eventual reward size.
Fishermen get their reward through a timely proof that at least one bonded party acted illegally. This will be especially valuable for detecting the ratification of invalid parachain blocks.


Fishermen are somewhat similar to “full nodes” in present-day blockchain systems that the resources needed are relatively small and the commitment of stable uptime and bandwidth is not necessary. Fishermen differ in so much as they must post a small bond. This bond prevents sybil attacks from wasting validators’ time and compute resources. 
While fishermen are part of the security model of Polkadot, the design would be secure without them. Since there is no incentive model for fishermen designed yet we need to keep Polkadot secure in their absence. Adding an incentive model for fishermen is part of our future work.

\begin{figure}[h]
	\centering
	\includegraphics[width=\textwidth]{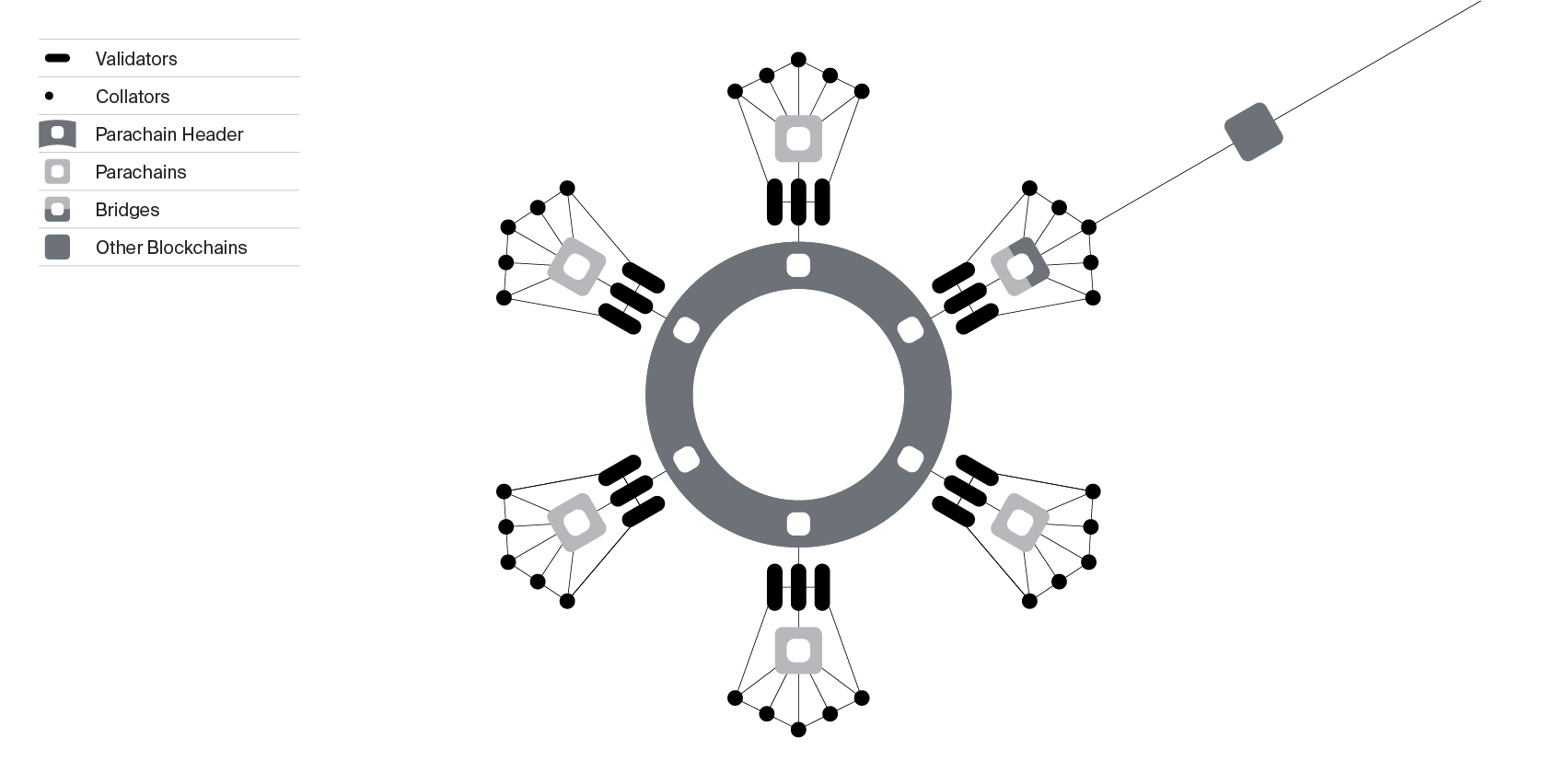}
	\caption{This figure shows a relay chain block securing six parachain blocks. Each parachain has  5 collators and 3 validators assigned to it (image credit: Ignasi Albero).}
	\label{fig:roles}
\end{figure}

\begin{figure}[h]
	\centering
	\includegraphics[width=\textwidth]{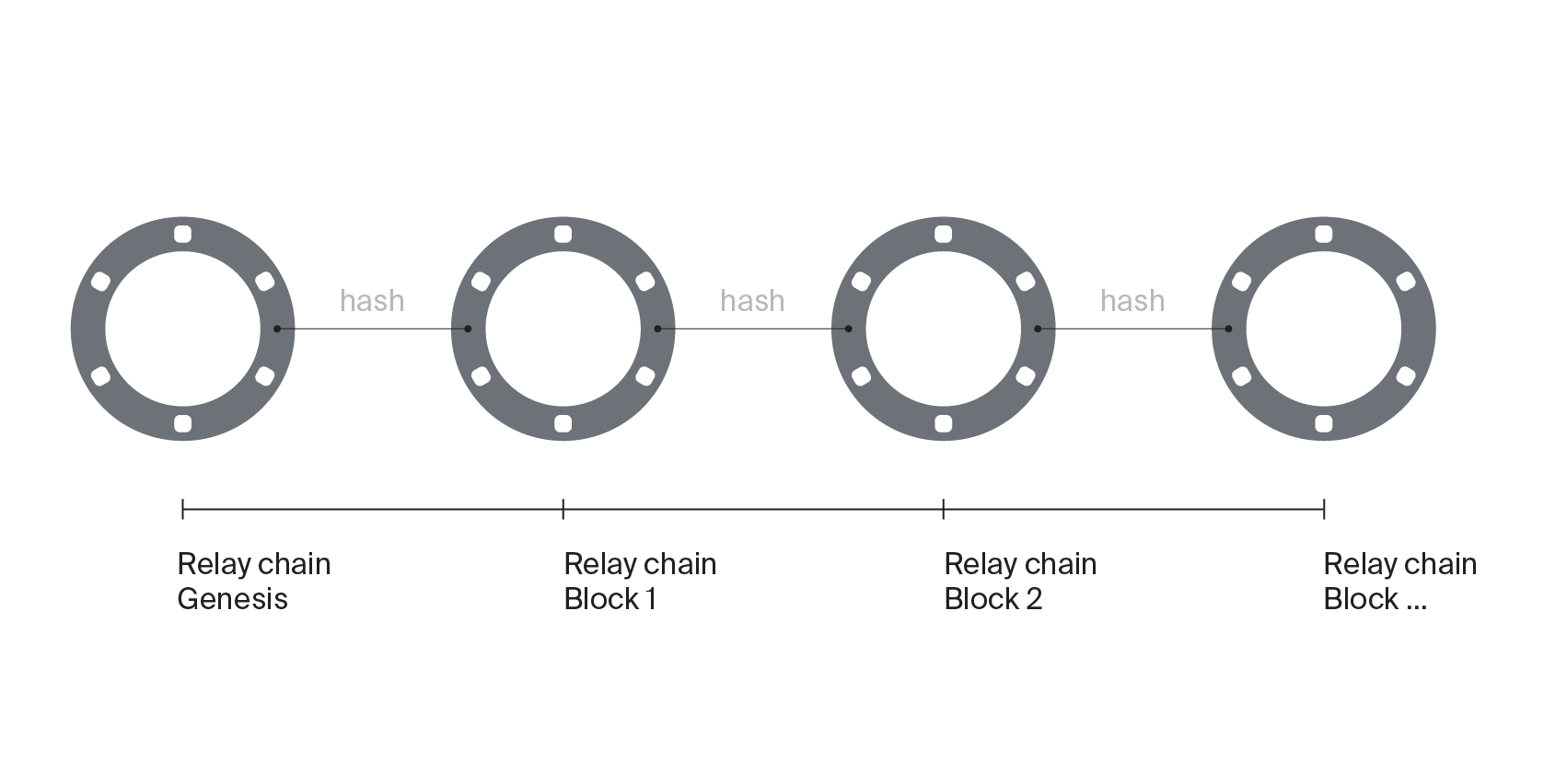}
	\caption{This figure shows the relay chain with 5 relay chain blocks. For simplicity, in this figure each relay chain block has parachain blocks from 6 parachains, however, the number of parachain blocks per relay chain block might vary (image credit: Ignasi Albero).}
	\label{fig:relaychain}
\end{figure}

The structural elements and different roles defined in the Polkadot protocol are shown in Figure~\ref{fig:roles}, in an example with six parachains, 18 validators, and 5 collators per parachain. Figure \ref{fig:relaychain} shows the relay chain with 5 such relay chain blocks. Note that the number of parachain validators assigned to a parachain is divided by the number of parachains, however, the number of collators is individual to parachains.  The bridge is a sub-protocol that allows external chains to interoperate with Polkadot, see \ref{sec:bridge} for more information. 

\subsection{Adversarial Model of Polkadot}\label{sec:security_model}

\paragraph{Roles:} In general, we assume that honest parties follow the protocol while malicious ones can follow any arbitrary algorithm. We assume that three quarters of nominators' stake belong to honest ones. As a result of this assumption, more than two third of validators who are elected by nominators are honest. We do not have any limit on number of malicious fishermen since their malicious behaviours are detectable and punishable.

\paragraph{Parachains:} We do not have any security assumption on block production mechanism for parachains. On the other hand, we assume that a significant amount of collators are honest. The security of Polkadot does not depend on any precise honest fraction of collators but it requires existence of some honest collators.

Parts of the protocol assume that every parachain has at least one reachable honest member; where this is not feasible or not realistic, we do not follow through on this assumption and instead have additional checks against a totally-malicious membership.

\paragraph{Keys:} We assume that malicious parties generate their keys with an arbitrary algorithm
while honest ones always generate their keys securely.

\paragraph{Network and Communication:} \label{par:network_coms} All validators have their own local clock and their clocks do not rely on any central clock.
We assume that  validators and collators are in a partially synchronous network.
It means that a message sent by a validator or a collator arrives at all parties in the network
at most $\D$ units of time later where $\D$ is an unknown parameter. So, we assume an eventual delivery of a message in Polkadot.
We also assume that collators and fishermen can connect to the relay chain network to submit their reports.

\section{Components and sub-protocols}\label{sec:components}
Next, we summarise Polkadot functionality shortly for an overall picture and then continue to describe the individual sub-protocols and components. 

Polkadot's validators are selected by the NPoS scheme (Section~\ref{sec:validators}). Nominated Proof-of-Stake or NPoS is our adaptation of PoS where an unlimited amount of token holders can participate as nominators, backing with their stake a large but limited set of validators. This paradigm simultaneously achieves high levels of security and scalability, as well as an unprecedented level of decentralisation by ensuring a property known in voting theory as proportional justified representation \cite{sanchez2017proportional, brill2017phragmen}. Nominators, who are economically vested in the security of the system, act as watchdogs over the validators' performance. Based on the nominators' expressed preferences over candidates, every era the system selects a set of validators with stake backings that are as high and as evenly distributed as possible. Nominators are also economically disincentivized from concentrating their votes on too few validators, which helps keep the system decentralised over time. Furthermore, the election mechanism is highly adaptive to sudden changes, such as some validators being kicked out after a slashing, as it automatically redistributes the nominators' backings across the new set of validators, even when the votes themselves do not change.

The security goal of Polkadot is to be Byzantine fault tolerant when the participants are rational (see Section \ref{sec:economics} for more details on incentives and economics). We assume that with the properties NPoS gives, the stakeholders elect a set of validators that has a more than $2/3$ fraction of honest members.

The elected validators are responsible for running the relay chain (Section~\ref{sec:relaychain}). While each parachain's collators are responsible for generating parachain blocks (Section~\ref{sec:parachainblockproduction}), the validators are divided into rotating subsets, one for each parachain, and need to attest to the validity of parachain blocks before the headers of those blocks are included in the relay chain.

To achieve good scalability the number of validators in each of these subsets is small. Nonetheless, thanks to the NPoS guarantee that every validator is well backed, the availability and validity scheme (Section~\ref{sec:validity-and-availability}) can ensure that any attack on the validity of Polkadot is very expensive in expectation. In fact, the entirety of Polkadot's economic security backs every parachain. This is in stark contrast to having, say, 100 independent blockchains with an equivalent sum total of stake, where on average each blockchain is backed by 1/100-th of the stake, and thus only benefits from 1/100-th the level of security. We guarantee availability by using erasure coding of each parachain block to make the validators collectively and robustly responsible for the availability of these blocks without breaking scalability.

For this to work, we need to be able to revert the chain until we know with good probability that all parachains are correct. This means that we need to be able to reorganise the chain and for that the chain needs to be capable of forking. Thus we use a block production mechanism, BABE (Section~\ref{sec:babe}), that while run by validators, has similar properties to proof-of-work chains. Specifically, we can use the longest chain rule as part of our consensus, and the next block producer is not known in advance. On its own BABE would require us to wait a long time from the moment a block is produced to the moment it is finalised, i.e.~when we can be confident that with high probability the block will never be reverted. Slow finality is required in some circumstances to deal with challenges to availability. Most of the time, however, we would prefer to finalise blocks much faster.  For this purpose, validators finalise blocks using GRANDPA (Section~\ref{sec:consensus}), a finality gadget that is cleanly separated from block production. This separation makes it very adaptive and here allows us to delay finalising blocks until challenges are dealt with, without slowing down block production. GRANDPA gets Byzantine agreement on finalised blocks and will allow us to prove to an entity that keeps track of the validator set which blocks are finalised, which will be important for bridges (Appendix~\ref{sec:bridge}).

If an account on one parachain sends tokens to another parachain, then XCMP (Section~\ref{sec:XCMP}) ensures that this message is delivered correctly. It is sent at a speed which is not dependent on how long it takes to finalise blocks, which means that it needs to deal with the possibility of Polkadot forking. Thus we optimistically execute based on the assumption that the parachain blocks are correct. If one is not, then we need to revert and for that, it is important that parachains only receive messages that were sent by blocks recorded on this new relay chain fork, and not the reverted fork. Thus we need that  the parachain and XCMP logic ensure that a fork of the relay chain defines a consistent history of Polkadot and thus messages only arrive when they have been sent previously in the history defined by this fork. 

If the token transfer is carried out in conjunction with SPREE modules (Appendix~\ref{sec:SPREE}) then that ensures that so long as the parachains execute correctly, tokens can only be created and destroyed in an agreed upon way. In turn the correct execution of the chains code is guaranteed by the availability and validity scheme. SPREE ensures that shared code needed for the token transfer logic is correct as well. Even though chains can change their own code, they will not be able to change the code of SPREE modules. Instead the code of SPREE modules is stored centrally and the execution of that code and its storage will be sandboxed from the rest of the state transition. This ensures that this token transfer message is interpreted correctly and obtains the guarantees about tokens we want. 

On the side of economics (Section~\ref{sec:economics}), we aim to have a controlled near-constant yearly inflation rate. As stated before, it is important for the security of the system that all validators have large amounts of stake backing them. Our adaptive reward schedule for validators and the nominators backing them ensures that overall participation in NPoS stays high, and that the validators' stake backings are evenly distributed. On a more granular level, we pay or slash validators on a per-executed-action basis, and extend the same rewards or punishment onto nominators proportionally, to ensure that the rational strategy is compatible with honest behaviour.

The relay chain's logic itself will need updating occasionally. The governance mechanism (Section~\ref{sec:governance}) allows Polkadot token holders to participate in the decision-making process rather than having any changes to the system be imposed by a central authority -- or in the case of some decentralised systems, by a team of developers. Too often, a contentious code change has led existing blockchains to an impasse or a permanent fork. 
We want a mechanism that balances being able to make uncontentious changes quickly when needed, while also providing the tools to deal with contentious proposals in a decisive and fair manner. The ultimate arbiters of Polkadot are the Dot token holders and so all important decisions, such as code changes, are made by state-weighted referenda. There is an elected council, responsible for making smaller decisions and partially setting the priority for referenda, in such a way that they cannot block a change that a majority wants.

Lastly we review 
some of the primitives that Polkadot sub-protocols are using such as the cryptographic keys and networking scheme in Section \ref{sec:crypto} and Section \ref{sec:networking}, respectively. Polkadot's networking needs to extend the peer-to-peer gossip network that is standard in single chain permissionless blockchains to a multi-chain system, where any nodes network traffic should not scale with the total data of the system.



\subsection{Nominated proof-of-stake and validator election}\label{sec:validators}
Polkadot will have a native token called DOT. It will use Nominated Proof-of-Stake (NPoS), our very own version of proof-of-stake (PoS).
Consensus protocols with deterministic finality, such as the one in Polkadot, 
require a set of registered validators of bounded size.
Polkadot will maintain a number $\nval$ of validators, in the order of hundreds or thousands.
This number will be ultimately decided by governance, and is intended to grow linearly with the number of parachains;
yet it will be independent of the number of users in the network, thus ensuring scalability.
However, NPoS allows for an unlimited number of DOT holders to participate as \emph{nominators},
who help maintain high levels of security by putting more value at stake.
As such, NPoS is not only much more efficient than proof-of-work (PoW),
but also considerably more secure than conventional forms of PoS such as DPoS and BPoS. 
Furthermore, we introduce new guarantees on decentralisation hitherto unmatched by any other PoS-based blockchain.

A new set of validators is elected at the beginning of every \emph{era} -- a period during roughly one day (see Table~\ref{t:time} in the Appendix) --
to serve for that era, according to the nominators' preferences.
More precisely, any DOT holder may choose to become a validator candidate or a nominator.
Each candidate indicates the amount of stake he is willing to stake and his desired commission fee for operational costs.
In turn, each nominator locks some stake and publishes a list with any number of candidates that she trusts.
Then a public protocol, discussed below, takes these lists as input and elects the candidates
with the most backing to serve as validators for the next era.

Nominators share the rewards, or eventual slashings, with the validators they nominated on a per-staked-dot basis; 
see Section~\ref{sec:economics} for more details. 
Nominators are thus economically incentivised to act as watchdogs for the system, and they will base their preferences 
on parameters such as validators' staking levels, commission fees, past performance, and security practices.
Our scheme allows for the system to elect validators with massive amounts of aggregate stake
- much higher than any single party's DOT holdings -
and thus helps turn the validator election process into a meritocracy rather than a plutocracy.
In fact, at any given moment we expect there to be a considerable fraction of all the DOT supply be staked in NPoS.
This makes it very difficult for an adversarial entity to get validators elected since it either needs a large amount of DOTs or high enough reputation to get the required nominators' backing,
as well as being very costly to attack because it is liable to lose all of its stake and its earned reputation.

Polkadot elects validators via a decentralised protocol with carefully selected, simple and publicly known rules,
taking the nominators' lists of trusted candidates as input. Formally, the protocol solves a multi-winner election
problem based on approval ballots, where nominators have voting power proportional to their stake,
and where the goals are decentralisation and security.

\paragraph{Decentralisation:}  
Our decentralisation objective translates into the classical notion of proportional representation in voting theory.
That is, a committee should represent each minority in the electorate proportional to their aggregate vote strength (in this case, their stake), with no minority being under-represented. 
We highlight here that nominators -- and their lists of trusted candidates -- constitute a valuable gauge for the preferences of the general community, and that diverse preferences and factions will naturally arise not only due to economical and security-related reasons, but also political, geographical, etc. Such diversity of points of view is expected and welcome in a decentralised community, and it is important to engage all minorities in decision-making processes to ensure user satisfaction. 

The goal of designing an electoral system that achieves proportional representation has been present in the literature for a very long time. Of special note is the work of Scandinavian mathematicians Edvard Phragm\'{e}n and Thorvald Thiele in the late nineteenth century. Very recently, there has been considerable effort in the research community to formalise the notion of proportional representation, and revisit the methods by Phragm\'{e}n and Thiele and optimise them algorithmically. 
Our validator selection protocol is an adaptation of Phragm\'{e}n's methods and is guaranteed 
to observe the technical property of \emph{proportional justified representation} (PJR)~\cite{sanchez2017proportional, brill2017phragmen}. 
Formally, this means that if each nominator $\nom \in \Nom$ has stake $stake_\nom$ 
and backs a subset $\Can_\nom\subseteq \Can$ of candidates,%
\footnote{For ease of presentation, we consider here a model where candidates have no stake of their own. 
	The general case can be reduced to this model by representing each candidate's stake as an additional nominator 
	that exclusively nominates that candidate.} %
the protocol will elect a set $\Val\subseteq \Can$ of $\nval$ validators such that, 
if there is a minority $\Nom'\subseteq \Nom$ of nominators such that %
$$|\cap_{\nom\in \Nom'} \Can_n| \geq t \quad \text{ and } \quad
\frac{1}{t} \sum_{\nom\in \Nom'} stake_\nom \geq \frac{1}{\nval} \sum_{\nom\in \Nom} stake_\nom,$$
for some $1\leq t\leq \nval$, then $|\Val\cap (\cup_{\nom\in\Nom'} \Can_n)| \geq t$.
In words, if a minority $\Nom'$ has at least $t$ commonly trusted candidates, 
to whom it could "afford" to provide with an average support of at least 
$\frac{1}{\nval} \sum_{\nom\in \Nom} stake_\nom$ 
(which in turn is an upper bound on the average validator support in the elected set $\Val$), 
then this minority has a justified claim to be represented in $\Val$ by at least $t$ candidates,  
though not necessarily commonly trusted.

\paragraph{Security:} If a nominator gets two or more of its trusted candidates elected as validators,
the protocol must also establish how to split her stake and assign these fractions to them.
In turn, these assignations define the total stake support that each validator receives.
Our objective is to make these validators' supports as high and as balanced as possible.
In particular, we focus on maximising the \emph{minimum validator support}.
Intuitively, the minimum support corresponds to a lower bound on the cost for an adversary to gain control
over one validator, as well as a lower bound on the slashable amount for a misconduct.

Formally, if each nominator $\nom \in \Nom$ has $stake_\nom$ and backs a candidate subset $\Can_\nom\subseteq \Can$,
the protocol must not only elect a set $\Val\subseteq \Can$ of $\nval$ validators
with the PJR property, but also define a distribution of each nominator's stake among the elected validators that she backs,
i.e. a function $f:\Nom\times\Val \rightarrow \mathbb{R}_{\geq 0}$ so that
$$\sum_{\val\in \Val\cap \Can_\nom} f(\nom, \val) = stake_\nom \quad \text{ for each nominator } \nom\in\Nom,$$
and the objective is then
$$\max_{(\Val, f)} \: \min_{\val\in \Val} \: support_f(\val),
\quad \text{ where } support_f(\val) := \sum_{\nom\in \Nom: \ \val\in \Can_\nom} f(\nom, \val). $$

The problem defined by this objective is called \emph{maximin support} in the literature~\cite{sanchez2016maximin}, and is known to be NP-hard.
We have developed for it several efficient algorithms which offer theoretical guarantees 
(constant-factor approximations), and also scale well and have been successfully tested on our testnet. 
For more information, see our paper on validator election in NPoS~\cite{NPoSpaper}.

\subsection{Relay Chain State Machine}\label{sec:relaychain}

Formally, Polkadot is a replicated sharded state machine where shards are the parachains and the Polkadot relay chain is part of the protocol ensuring global consensus among all the parachains. Therefore, the Polkadot relay chain protocol, can itself be considered as a replicated state machine on its own. In this sense, this section describes the relay chain protocol by specifying the state machine governing the relay chain. To that end, we describe the relay chain state and the detail of state transition governed by transactions grouped in the relay chain blocks.

\paragraph{State:} The state is represented through the use of an \emph{associative array} data structure composed by a collection of $(key, value)$ pairs where each key is unique. There is no assumption on the format of the key or the value stored under it besides the fact that they both the key and the value need to be finite byte arrays.

The $(key, value)$ pairs which comprise the relay chain state are arranged in a Merkle radix-16 tree. The root of this tree canonically identifies the current state of the relay chain. The Merkle tree also provides an efficient mean to produce the  proof of inclusion for an individual pair in the state.

To keep the state size in control, the relay chain state is solely used to facilitate the relay chain operations such as staking and identifying validators. The Merkle Radix tree is not supposed to store any information regarding the internal operation of the parachains.

\paragraph{State transition: } \label{par:state_transition} Like any transaction-based transition system, Polkadot state changes via an executing ordered set of instructions, known as extrinsics. These extrinsics include transactions submitted by the public. They cover any data provided from ``outside'' of the machine's state which can affect state transition. Polkadot relay chain is divided into two major components, namely the ``Runtime'' and the ``Runtime environment''. The execution logic of the state-transition function is mainly encapsulated in the Runtime while all other generic operations, commonly shared among modern blockchain-based replicated state machines, are embedded into the Runtime environment. In particular, the latter is in charge of network communication, block production and consensus engines.

Runtime functions are compiled into a WebAssembly module and are stored as part of the state. The Runtime environment communicates the extrinsics to the Runtime and interacts with it to execute the state transition. In this way, the state transition logic itself can be upgraded as a part of the state transition.

\paragraph{Extrinsics:} \label{par:extrinsics}

Extrinsics are the input data supplied to the Polkadot relay-chain state machine to transition to new states. Extrinsics need to be stored into blocks of the relay chain in order to achieve consensus among the state machine replica. Extrinsics are divided into two broad categories namely: transactions and "inherents" which represent data that is inherent to a relay chain block. The timestamp $t$ of a block is an example of inherent extrinsics which must be included in each Polkadot relay chain block.

Transactions are signed and are gossiped around on the network between nodes. In contrast, inherents are not signed and are not gossiped individually but rather only when they are included in a block. The inherents in a block are assumed to be valid if a supermajority of validators assumes so.  
Transactions on the relay chain are mainly concerned with the operation of the relay chain and Polkadot protocol as a whole, such as \texttt{set\_code}, \texttt{transfer}, \texttt{bond}, \texttt{validate}, \texttt{nominate}, \texttt{vote}.

Relay chain block producers listen to all transaction network messages. Upon receiving a transaction message, the transaction(s) are validated by the Runtime. The valid transactions then are arranged in a queue based on their priority and dependency and are considered for inclusion in future blocks accordingly.

\paragraph{Relay chain block format:}
A typical relay chain block consists of a header and a body. The body simply consists of a list of extrinsics.

The header contains the \textit{hash of parent block}, \textit{block number}, the \textit{root of the state tree}, the \textit{root of the Merkle tree} resulting from arranging the extrinsics in such a tree and the \textit{digest}. The digest stores auxiliary information from the consensus engines which are required to validate the block and its origin as well as information helping light clients to validate the block without having access to the state storage.

\paragraph{Relay chain block building:}\label{sec:relaychainblockproduction}
In this section, we present a summary of various steps of relay chain operation which are carried out by its validators. A priori, each validator privately knows the times during which it is supposed to produce a block (see \ref{sec:babe}).

Meanwhile, transactions ranging from the validated parachain block hash, transfer, staking, nomination or slashing for protocol violation are submitted to the relay chain validators. The validators examine the validity of the transactions and store them in their transaction pool. Once the time slot during which the validator is expected to produce the block has arrived, the validator estimates the block which most likely represents the state which is going to be finalised by the finality protocol and set it as the current state of the relay chain. Then the validator selects valid transactions with from the transaction pool, executes them and updates the state accordingly. The validator executes and collates as many transactions as the block capacity allows and attaches a cryptographic digest of the final stage of the relay chain after executing the selected transactions. Finally the validator signs and publishes the built block.

Upon receiving the new block, other validators examine the producer's adherence to the protocol as well as the validity of included transactions and store the block in the \emph{block tree} which represents all possible candidates for a final state transition of the relay chain. 

Simultaneously, the set of validators votes on various branches of the block tree (see \ref{sec:grandpa}) and prunes branches which conflict with the version agreed upon by the supermajority of the validators. In that way, they eventually agree on a canonical state of the relay chain.

\subsection{Consensus}\label{sec:consensus}

In this section, we explain the hybrid consensus protocol of Polkadot which consists of BABE: a block production mechanism of the relay chain that provides probabilistic finality and GRANDPA which provides provable, deterministic finality and works independently from BABE.  Informally probabilistic finality implies that after certain time passes, a block in the relay chain will be finalised with very high probability (close to 1) and deterministic finality implies a finalised block stays final forever. Furthermore provable finality means that we can prove to parties not actively involved in the consensus that a block is final.

We need provable finality to make bridges to chains outside Polkadot easier; another blockchain, not part of Polkadot's consensus could then be convinced of when it is safe to act on data in a relay chain or parachain block without any danger of it being reverted. The best way of getting that is to have Byzantine agreement among the validators on the state of Polkadot and its parachains. However the availability and validity scheme \ref{sec:validity-and-availability} may also require us to revert blocks, which would mean that getting Byzantine agreement on every block, as in Tendermint \cite{Tendermint} or Algorand \cite{ALGORAND}, would not be suitable. However, this should happen rarely as a lot of stake will be slashed when we do this. As a result, we want a scheme that generates blocks and optimistically executes them, but it may take some time to finalise them. Thus GRANDPA voters need to wait for assurances of availability and validity of a block before voting to finalise that block.
Even the speed at which we finalise blocks may vary - if we do not receive reports of invalidity and unavailability then we can finalise fast, but if we do then we may need to delay finality while we execute more involved checks. 

Because of the way Polkadot's messaging protocol (XCMP \ref{sec:XCMP}) works, message passing speed is constrained by block time, but not by finality time. Thus if we delay finality but in the end do not revert, then message passing is still fast.

As a result of these requirements, we have chosen to separate the mechanisms for block production and finalising blocks as much as possible. In the next two sections, we describe the protocols BABE and GRANDPA that do these respectively.

\subsubsection{Blind Assignment for Blockchain Extension (BABE)}
\label{sec:babe}

In Polkadot, we produce relay chain blocks using our Blind Assignment for Blockchain Extension protocol (BABE).
BABE assigns validators randomly to block production slots using  the randomness generated with blocks. A block production slot is a division of time when a block producer may produce a block. Note, that time is not universally agreed on, which we will address later.  These assignments are completely private until the assigned validators produce their blocks. Therefore, we use ``Blind Assignment'' in the protocol name. BABE is similar to Ouroboros Praos \cite{Praos} with some significant differences in the chain selection rule and timing assumptions.

In BABE, we may have slots without any assignment
which we call empty slot. 
In order to fill the empty slots, we have a
secondary block production mechanism based on Aura \cite{aura} that assigns validators to slots publicly. We note that these blocks do not contribute to
the security of BABE since the best chain selection and the random number generation algorithms work as if Aura blocks do not exist.
Therefore, next we only describe BABE together with its security properties.

BABE \cite{babe} consists of another time division called \emph{epochs} ($e_1,e_2,...$), where each epoch consists of a number of sequential block production slots (\(e_i = \{sl^i_{1}, sl^i_{2},\ldots,sl^i_{t}\}\)) up to the bound  $R$.
Each validator knows in which slots it is supposed to produce a block at the beginning of every epoch. When the time for its slot comes, the validator produces the block by proving that it is assigned to
this slot.

The blind assignment is based on the cryptographic primitive called verifiable random function (VRF) \cite{vrf} (see Section \ref{sec:session_keys}). 
A validator in an epoch $e_m$  does the following to learn if it is eligible to produce a block in slot $sl_i^m$:
\begin{enumerate}
	\item  it obtains the randomness in the genesis block if $ m = 1  $ or $ m =2 $. Otherwise, it obtains the randomness  generated two epochs before ($e_{m-2}$).
	\item  it runs the VRF with its secret key and the input:  randomness and the slot number $ sl_i^m $.
\end{enumerate}

If the output of  VRF is less than the threshold $ \tau $, then the validator is the slot leader meaning that it is eligible to produce a block for this slot. We select $\tau$ with respect to security requirements of BABE \cite{babe} e.g., bigger $ \tau $ makes less probable to select only honest validators for a slot than smaller $ \tau $. 
When a validator produces a block, it adds the output of the VRF and its proof to the block which shows that its VRF output is less than $\tau$  in order to convince other validators that it has a right to produce a block in the corresponding slot. The validators always generate their blocks on top of the best chain.
The best chain selection rule in BABE says that ignore the Aura blocks and select the longest chain that includes the last finalised GRANDPA block. See Section \ref{sec:grandpa} for the details how blocks are finalised in GRANDPA.

The randomness of an epoch $e_m$ where $ m > 2 $ is generated by using the BABE blocks of the best chain that belongs to that epoch: let \(\rho\) be the concatenation of all  VRF values in BABE blocks that belongs to $e_m$. Then, compute the randomness for epoch $ e_m $ as $r_{m} = H(m
||\rho)$ where $ H $ is a hash function. Validators run periodically the relative time algorithm described below to learn at what time a slot starts according to their local clocks.

\paragraph{Relative Time Protocol:}

The elected validators for a slot need to know when the right time is to produce a block for the consistency and the security of BABE. For this, validators uses their local computer clock which is not adjusted by  any centralized clock adjustment protocols such as the Network Time Protocol \cite{ntp}. Instead, they keep their clock synchronised with the other validators with the relative time protocol. 
The formal security model of local clock synchronisation in blockchains without NTP and further details about the relative time protocol can be found in \cite{consensusonclock}.

In BABE, we assume that after the genesis block is released, elected validators of the first epoch store the arrival time of the genesis block with respect to their local clock. Then, they mark the beginning
time of the first slot and increment
the slot number every $ T $ seconds. After this point,  they periodically run the relative algorithm not to lose the synchronisation with others because of their local clock drifts.  In addition to this, a validator who
joins after the genesis block runs the relative time algorithm to be synchronised with the other validators.

\begin{figure}[h]
	\centering
	\includegraphics[width=1.\textwidth]{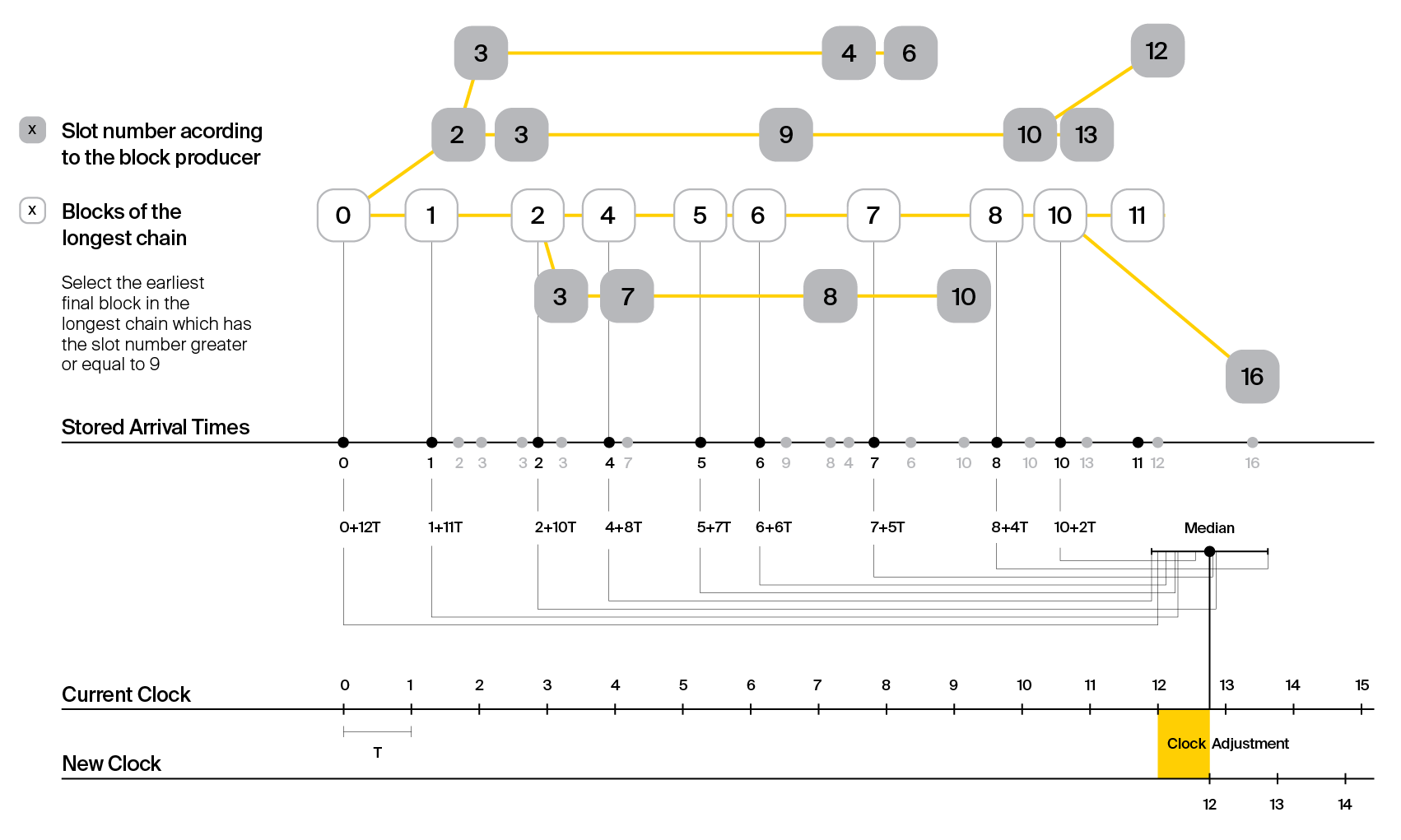}
	\caption{An example execution of the relative time protocol in the first epoch where $s_{cd} = 9$ (image credit: Ignasi Albero).}
	\label{fig:relativetime}
\end{figure}

In every sync-epochs (different than epochs in BABE), validators update their clock according to the result of the relative time protocol and use the new clock until the next sync-epoch. The first sync-epoch $\varepsilon_1$ starts just after the genesis block is released. The other sync-epochs  $\varepsilon_i$ start when the slot number of the last (probabilistically) finalised block is $\bar{sl}_{\epsilon}$ which is the smallest slot number such that  $\bar{sl}_{\varepsilon} - \bar{sl}_{\varepsilon-1} \geq s_{cd}$ where $\bar{sl}_{\varepsilon-1}$ is the slot number of the last (probabilistically) finalized block in sync-epoch $\varepsilon-1$. Here, $s_{cd}$ is the parameter of the chain density (CD) property which will be defined according the chain growth.
In more detail, each validator  stores  the arrival time $ t_j $ of  blocks together with the slot number $\slot'_j$ in the block during a sync-epoch. At the end of a sync-epoch, the validator retrieves the arrival time of probabilistically finalized blocks generated during the sync-epoch and computes some candidate start times of the first slot $ \slot $ of the next sync-epoch i.e,  given that $ a_j = T(\slot - \slot'_j)  $,  $\mathcal{C}_T = \{t_j+a_j \}$. The times in $ \mathcal{C}_T $ are considered as candidates. In order to  choose one candidate,  the validator then sorts the list of candidates $ \mathcal{C}_T $ and outputs the median of the sorted list as a start time of the $ \slot $. An example execution of the relative time protocol in the first sync-epoch is in Figure \ref{fig:relativetime}.

\paragraph{Security Overview of BABE:} Garay et al. \cite{backbone} define the properties below in order to obtain a secure blockchain protocol. Informally, we can describe these properties as follows:

\begin{itemize}
	\item \emph{Common Prefix (CP):} \label{item:common_prefix}
	It ensures that the blocks which are $ k $-blocks before the last block of an honest validator's blockchain cannot be changed. We call  all unchangeable blocks  \emph{finalized} blocks. BABE satisfies CP property thanks to the honest super majority since malicious validators are selected for a slot probabilistically much less than the honest validators. It means that malicious validators do
	not have enough 
	to construct another chain which does not include one of the finalised blocks.
	\item \emph{Chain Quality (CQ):} \label{item:chain_quality} It ensures a minimum honest block contribution to any best chain owned by an honest party in every certain number of slots. We guarantee even in the worst case where a network delay is maximum that there will be at least one honest block in the best chain during an epoch so that the randomness cannot be biased.
	\item \emph{Chain Growth (CG):} \label{item:chain_growth} It guarantees a minimum growth between slots. Thanks to super majority of honest validators, malicious validators cannot prevent the growth of the best chain.
	
	\item \emph{Chain Density (CD):} \label{item:chain_density} It ensures that in a sufficiently long portion of the best chain more than half of the blocks produced by honest validators. CQ and CG properties 
	imply this property \cite{Praos}.
\end{itemize}
Further details about BABE and  its security analysis can be found in \cite{babe}.


\subsubsection{GRANDPA} \label{sec:grandpa}

As mentioned above, we want a finalisation mechanism that is flexible and separated from block production, which is achieved by GRANDPA. The only modification to BABE required for it to work with GRANDPA is to change the fork-choice rule: instead of building on the longest chain, a validator producing a block should build on the longest chain including all blocks that it sees as finalised. GRANDPA can work with many different block production mechanisms and it will be possible to switch out BABE with another.

Intuitively GRANDPA is a Byzantine agreement protocol that works to agree on a chain, out of many possible forks, by following some simpler fork choice rule, which together with the block production mechanism would give probabilistic finality if GRANDPA itself stopped finalising blocks. We want to be able to agree on many blocks at once, in contrast to single-block Byzantine agreement protocols.

We assume that we can ask the fork choice rule for the best block given a particular block. The basic idea is that we want to reach Byzantine agreement on the prefix of the chain that everyone agrees on. To make this more robust, we try to agree on the prefix of the chain that 2/3 of validators agree on.

\begin{figure}[h!]
	\centering
	\includegraphics[width=0.7\textwidth]{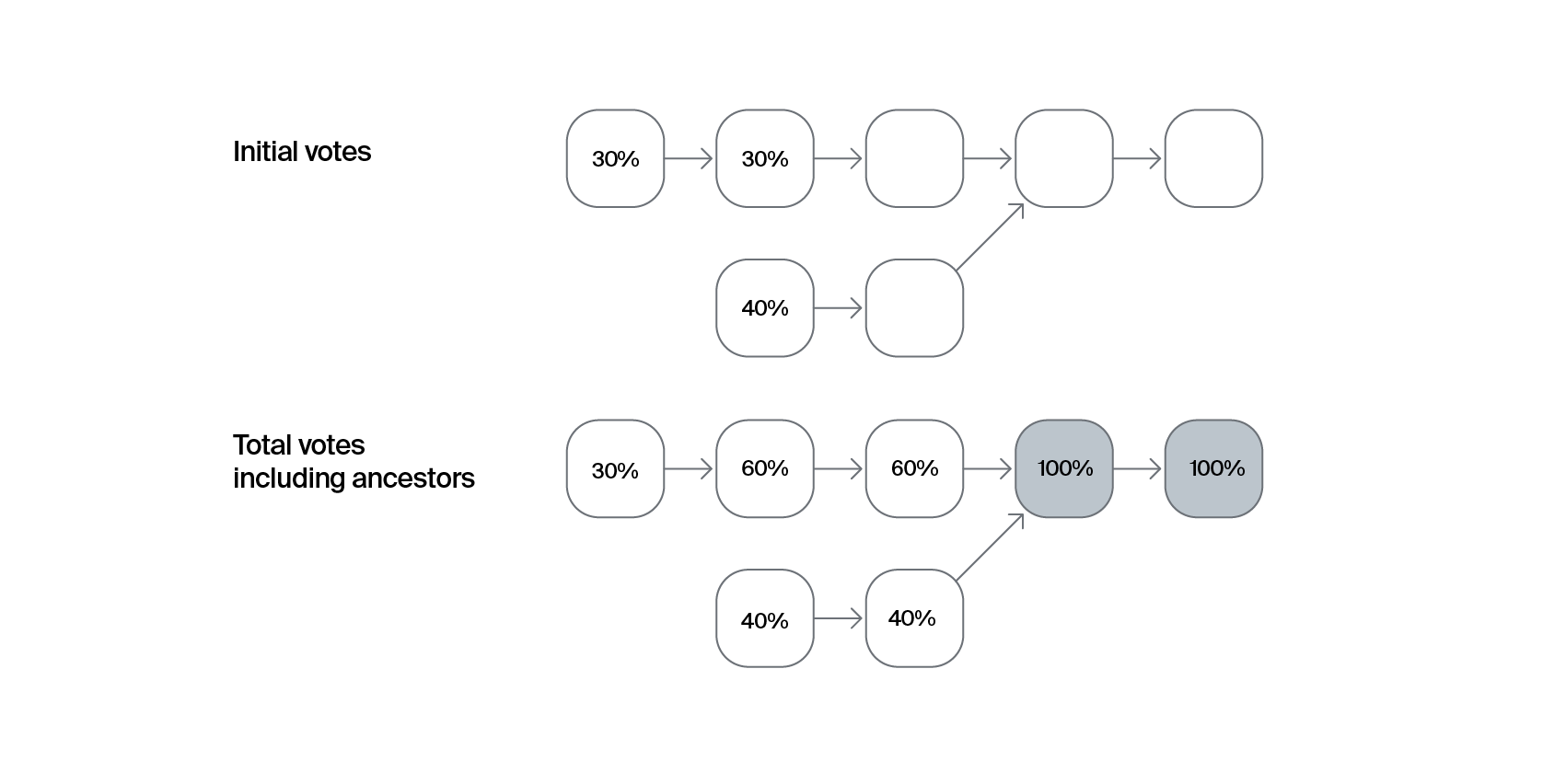}
	\caption{GRANDPA votes and how they are aggregated (image credit: Ignasi Albero).}
	\label{fig:grandpa}
\end{figure}

We make use of a Greedy Heaviest Observed Subtree (GHOST) on votes rule, much like Casper TFG \cite{CasperTFG} or some of the fork choice rules suggested for use with Casper FFG \cite{CasperFFG}. We use this rule inside what is structured like a more traditional Byzantine agreement protocol, to process votes. The 2/3 GHOST rule (pictured in Figure \ref{fig:grandpa})  works as follows. We have a set of votes, given by block hashes in which honest validators should not have more than one vote, and we take the head of the chain formed inductively as follows. We start with the genesis block and then include the child of that block that 2/3 of the voters voted for descendants of, as long as there is exactly one such child. The head of this chain is $g(V)$ where $V$ is the set of votes. For example, in Figure \ref{fig:grandpa}, the left hand side gives the votes for individual blocks and the right hand side the total votes for each block and all of its descendants. The genesis block is at the top and we take its child with 100\% $> 2/3$ of the votes. The children of that block have 60\% and 40\% of the votes respectively and since these are below $2/3$ we stop and return the second block.

There are two voting phases in a round of GRANDPA: prevote and precommit. Firstly validators prevote on a best chain. Then they apply the 2/3-GHOST rule, $g$, to the set of prevotes $V$ they see and precommit to $g(V)$. Then similarly they take the set of precommits $C$ they see and finalise $g(C)$.

To ensure safety, we ensure that all votes are descendants of any block that could possibly have been finalised in the last round. Nodes maintain an estimate of the last block that could have been finalised in a round, which is calculated from the prevotes and precommits. Before starting a new round, a node waits until it sees enough precommits for it to be sure that no block on a different chain or later on the same chain as this round's estimate can be finalised. Then it ensures that it only prevotes and precommits in the next round to blocks that are descendants of the last round's estimate which it keeps updating by listening to precommits from the last round. This ensures safety.

To ensure liveness, we select one validator in rotation to be the primary. They start the round by broadcasting their estimate for the last round. Then when validators prevote, if the primary's block passes two checks, that it is at least the validator's estimate and that it got $>2/3$ prevotes for it and its descendants in the last round, then it prevotes for the best chain including the primary's block. The idea here is that if the primary's block has not been finalised, then progress is made by finalising the block. If the primary's block has not been finalised and all validators agree on the best chain including the last finalised block, which we should do eventually because BABE gives probabilistic finality on its own, then we now make progress by finalising that chain.

\subsection{Parachains}\label{sec:parachains}
In this section we review parachain's block production, their availability and validity scheme, and their messaging scheme.
\subsubsection{Block Production}\label{sec:parachainblockproduction}

We will discuss block production for a general parachain. At the end of the section, we will discuss alternatives.

In outline, a collator produces a parachain block, sends it to the parachain validators,
who sign its header as valid, and the header with enough signatures is placed on the relay chain.
At this point, the parachain block is as canonical as the relay chain block its header appeared in.
If this relay chain block is in the best chain according to BABE (see Section \ref{sec:babe}), so is the parachain block and when this relay chain block is finalised, so is the parachain block.

Because the parachain validators switch parachains frequently, they are stateless clients of the parachain.
Thus we distinguish between the parachain block $B$, which is normally enough for full nodes of the parachain such as collators to update the parachain state,
and the {\em Proof of Validity(PoV)} block $B_{PoV}$, which a validator who does not have the parachain state can verify.

Any validator should be able to verify $B_{PoV}$ given the relay chain state using the parachain's {\em state transition validation function} (STVF),
the Web assembly code for which is stored on the relay chain in a similar way to the relay chain's runtime.
The STVF takes as an input the PoV block, the header of the last parachain block from this parachain and a small amount of data from the relay chain state.

The STVF outputs the validity of the block, the header of this block and its outgoing messages.
The PoV block contains any outgoing messages and the parachain block $B$. The parachain validators should gossip the parachain block to the parachain network, as a back up to the collator itself doing so.

The PoV block will be the parachain block, its outgoing messages, its header and light client proof witnesses. These witnesses are Merkle proofs that give all elements of the input and output state that are used or modified by the state transition from the input and output state roots.

To aid in censorship resistance, a parachain may want to use proof of work or proof of stake to select collators, where the selection strategy is up to the given parachain.
This can be implemented in the STVF and need not be a part of the Polkadot protocol. So for proof of work,
the STVF would check that the hash of the block is sufficiently small.
However, for speed, it would be useful to ensure that most relay chain blocks can include a parachain block.
For PoW, this would necessitate it being probable that multiple collators are allowed to produce a block.
As such we will still need a tie-breaker for the parachain validators to coordinate on validating the same parachain block first.
This may be the golden ticket scheme of \cite{2016:Wood:Polkadot}. For proof of stake this may not be necessary.

Optionally, for some parachains, the parachain block $B$ may not be enough for collators to update their state. This may happen for chains that use succinct zero-knowledge proofs to update their state, or even for permissioned chains that just give signatures from authorities for validity. Such chains may have another notion of parachain block which is actually needed to update their state and must have their own scheme to guarantee the availability of this data.

\subsubsection{Validity and Availability} \label{sec:validity-and-availability}
Once a parachain block is created it is important that the {\em parachain blob} consisting of the PoV block and set of outgoing messages from the parachain is available for a while.
The naive solution for this would be broadcasting/gossip the parachain blobs to all relay chain nodes, which is not a feasible option because there are many parachains and the PoV blocks may be big. We want to find an efficient solution to ensure PoV blocks from any recently created parachain blocks are available.

For a single chain, such as Bitcoin, as long as 51\% of hash power is honest, not making block data available ensures that no honest miner builds on it so it will not be in the final chain. However, parachain consensus in Polkadot is determined by relay chain consensus.
A parachain block is canonical when its header is in the relay chain.
We have no guarantees that anyone other than the collator and parachain validators have seen the PoV block.
If these collude then the rest of the parachain network need not have the parachain block and then most collators cannot build a new block and this block's invalidity may not be discovered. We would like the consensus participants, here the validators, to collectively guarantee the availability rather than relying on a few nodes.

To this end we designed an availability scheme that uses erasure coding (see e.g. \cite{availabilityETH2}) to distribute the PoV block to all validators.
When any misbehaviour, particularly in relation to invalidity, is detected, the blob can be reconstructed from the distributed erasure coded pieces.

If a block is available then full nodes of the parachain, and any light client that has the PoV block, can check its validity. We have three-level of validity checks in Polkadot. The first validity check of a PoV block is executed by the corresponding parachain validators. If they verify the PoV block then they sign and distribute the erasure codes of the blob, including the PoV block, to each validator.
We rely on nodes acting as fishermen to report the invalidity of a blob as  a second level of validity checking. They would need to back any claim with their own stake in DOTs. We would assume that most collators will be fishermen, as they have a stake in continued validity of the chain and are already running full nodes, so all they need is stake in DOTs. The third level of validity checking is executed by a few randomly and privately assigned validators. We determine the number of validators in the third level of validity checking considering the amount of invalidity reports given by fishermen and unavailability reports given by  collators. If an invalid parachain block is detected, the validators who signed for its validity are slashed.
We wait for enough of these randomly assigned checkers to check the block before voting on it in GRANDPA. We also  want to ensure that the block is available before selecting the randomly assigned validators. This means that the parachain validators have to commit running a high risk of being slashed for a small probability of getting an invalid block finalised. This means that the expected cost of getting an invalid block into Polkadot is higher than the amount of stake backing a single parachain.

The security of our availability and validity scheme is based on the security of the GRANDPA finality gadget (see Section \ref{sec:grandpa}) and the quality of randomness generated in each BABE epoch (see Section \ref{sec:babe}). Please see \cite{availandvalid} for more details about the availability and validity scheme.

\subsubsection{Cross-chain Message Passing (XCMP)} \label{sec:XCMP}
XCMP is the protocol that parachains use to send messages to each other.
It aims to guarantee four things: first that messages arrive quickly; second that messages from one parachain arrive to another in order; third that arriving messages were indeed sent in the finalised history of the sending chain; and fourth that recipients will receive messages fairly across senders, helping guarantee that senders never wait indefinitely for their messages to be seen.

There are two parts to XCMP. (1) Metadata about outgoing messages for a parachain block are included on the relay chain and later this metadata is used to authenticate messages by the receiving parachain. (2) The message bodies corresponding to this metadata need to be actually distributed from the senders to the recipients, together with a proof that the message body is actually associated with the relevant metadata. The details of distribution are covered as a networking protocol in \nameref{sec:net_crosschain}; the remainder is covered below.

The way relay chain blocks include headers of parachain blocks gives a synchronous notion of time for parachain blocks, just by relay chain block numbers. Additionally it allows us to authenticate messages as being sent in the history given by the relay chain i.e. it is impossible that one parachain sends a message, then reorgs \footnote{reorganisation of the chain} so that that message was not sent, but has been received. This holds even though the system may not have reached finality over whether the message was sent, because any relay chain provides a consistent history.

Because we require parachains to act on every message eventually, non-delivery of a single message can potentially stop a parachain from being able to build blocks. Consequently we need enough redundancy in our message delivery system. Any validators who validate the PoV block should keep any outgoing messages from that block available for a day or so and all full nodes of the sending parachain also store the outgoing messages until they know they have been acted on.

To achieve consistency, when a source parachain $S$ sends messages in a parachain block $B$ to a destination parachain $D$, then we will need to authenticate these using the relay chain state, which is updated based on the parachain header $PH$ corresponding to the parachain block $B$ that was included in the relay chain. We need to limit the amount of data in headers like $PH$, in the relay chain state and also to limit what the relay chain needs to do for authentication when processing such parachain headers.

To this end, the parachain header $PH$ contains a message root $M$ of outgoing messages, as well as a bitfield indicating which other parachains were sent messages in this block.
The message root $M$ is the root of a Merkle tree of the head hash $H_p$ of a hash chain for each parachain $p$ that this block sends messages to.  The hash chain with head $H_D$ has the hash of all messages sent to $S$ from $D$, not just in block $B$ but ever sent from $S$ to $D$ in any block up to block $B$. This allows many messages from $S$ to $D$  to be verified at once from $M$. However the messages themselves are passed, they should also be sent with the Merkle proof that allows nodes of the receiving parachain
to authenticate that they were sent by a $B$ whose header  $PH$ was in a particular relay chain block.

Parachains receive incoming messages in  order. Internally parachains may defer or reorder acting on messages according to their own logic (possibly constrained by SPREE, see \ref{sec:SPREE}). However they must receive messages in the order determined by the consistent history given by the relay chain.
A parachain $D$ always receives messages sent by parachain blocks whose header was in earlier relay chain blocks first. When several such source parachains have a header in the relay chain block, the messages from these parachains are received in some predetermined order of parachains, either sequentially in order of increasing parachain id, or some shuffled version of this.

A parachain $ D  $ receives all messages sent by one parachain $ S $ in one parachain block or none of them.
A parachain header $PH'$ of $D$ contains a watermark. This watermark consists of a block number of a relay chain block $R$ and a parachain id of a source parachain $S$. This indicates that $D$ has recieved all messages sent by all chains before relay chain block $R$, and has acted on messages sent in block $R$ from parachains up to and including $S$ in the ordering.

The watermark must advance by at least one sending parachain in each of $ D $’s parachain blocks, which means that the watermark's relay chain block number advances or it stays the same and we only advance the parachain. To produce a parachain block on parachain $D$ which builds on a particular relay chain block $R$, a collator would need to look at which parachain headers were built between the relay chain block that the last parachain block of this chain built on. In addition, it needs the corresponding message data for each of those that indicated that they sent messages to $D$.
Thus it can construct a PoV block so that the STVF can validate that all such messages were acted on. Since a parachain must accept all messages that are sent to it,
we implement a method for parachains to make it illegal for another parachain to send it any messages that can be used in the case of spam occurring. When the parachain header of a parachain block that sends a message is included in a relay chain block, then any nodes connected to both the source and destination parachain networks should forward messages, together with their proofs, from sender to receiver.
The relay chain should at least act as a back up: the receiving parachain validators  of $D$ are connected to $D$'s parachain network and if they do not receive messages on it, then they can ask for them from the parachain validators of the sending chain $S$ at the time the message was sent.

\subsection{Economics and Incentive Layer}\label{sec:economics}

Polkadot will have a native token called DOT. Its various functions are described in this section.

\subsubsection{Staking rewards and inflation}\label{sec:inflation}

We start with a description of staking rewards, i.e.~payments to \emph{stakers} -- validators and nominators -- 
coming from the minting of new DOTs. 
Unlike some other blockchain protocols, the amount of tokens in Polkadot will not be bounded by an absolute constant, but there will rather be a controlled yearly inflation rate. Indeed, recent research~\cite{chitra2019competitive} suggests that in a proof-of-stake based protocol the staking rewards must remain competitive, in order to maintain high staking rates and high security levels, so deflationary policies are advised against. 

In our design, staking rewards are the only mechanism that mints DOTs. 
Thus, it is convenient to introduce our inflation model in this section as well. 

Recall from the description of the NPoS protocol (Section \ref{sec:validators}) that both validators and nominators stake DOTs. 
They get paid roughly proportional to their stake, but can be slashed up to $100\%$ in case of a misconduct. 
Even though they are actively engaged for only one era%
\footnote{Recall that an era lasts approximately one day. See Table~\ref{t:time} in the Appendix.} 
at a time, they can continue to be engaged for an unlimited number of eras. 
During this period their stake is locked, meaning it cannot be spent, and it remains locked for several weeks after their last active era, to keep stakers liable to slashing even if an offence is detected late.

\paragraph{Staking rate, interest rate, inflation rate:} Let the staking rate be the total amount of DOTs 
currently staked by validators and nominators, divided by the current total DOT supply. 
The stakers' average interest rate will be a function of the staking rate: 
if the staking rate dips below a certain target value selected by governance, 
the average interest rate is increased, thus incentivising more participation in NPoS, and vice versa. 
For instance, a target staking rate of $50\%$ could be selected as a middle ground between security and liquidity. 
If the stakers' average yearly interest rate is then set to $20\%$ at that level, 
we can expect the inflation rate to fluctuate closely around $50\%\times 20\% = 10\%$. 
Hence, by setting targets for the staking rate and stakers' interest rate, we also control the inflation rate. 
Following this principle, every era we adjust our estimate of the staking rate, 
and use it to compute the total amount of DOTs to be paid to stakers for that era.

\paragraph{Rewards across validator supports:} 
Once the total payout for the current era is computed, we need to establish how it is distributed.
Recall that the validator election protocol (Section \ref{sec:validators}) partitions the active stake into 
\emph{validator supports}, where each validator support is composed of the full stake of one validator 
plus a fraction of the stake of its backing nominators, and this partition is made so as to make validator supports 
as high and evenly distributed as possible, hence ensuring security and decentralisation. 
A further incentive mechanism put in place to ensure decentralisation over time 
is paying validator supports equally for equal work, regardless of their stake. 
As a consequence, if a popular validator has a high support, its nominators will likely be paid less per staked DOT 
than nominators backing a less popular validator. Hence, nominators will be incentivised to change their preferences 
over time in favour of less popular validators (with good reputation nonetheless), helping the system converge to the ideal case where all validator supports have equal stake.

In particular, we devise a point system in which validators accumulate points for each payable action performed, 
and at the end of each era validator slots are rewarded proportional to their points. 
This ensures that validators are always incentivised to maintain high performance and responsiveness. 
Payable actions in Polkadot include: a) validating a parachain block, 
b) producing a relay chain block in BABE, 
c) adding to a BABE block a reference to a previously unreferenced uncle block,%
\footnote{In the BABE protocol, at times two block producers may generate different blocks A and B at the same height, leading to a temporary fork in the relay chain. The fork will quickly be resolved and one of the blocks selected, say A, as part of the main chain, while block B becomes an \emph{uncle} to all descendents of A. For security reasons, it is convenient to record and timestamp \emph{all} blocks produced, but since uncle blocks cannot be accessed via parent relations, we encourage block producers to explicitly add these references to the main chain.}
and d) producing an uncle block.

\paragraph{Rewards within a validator slot:} As a nominator's stake is typically split among several validator supports, 
their payout in an era corresponds to the sum of their payouts relative to each of these supports. 
Within a validator support, the payment is as follows: 
First, the validator is paid a \emph{commission fee}, which is an amount intended to cover its operational costs. 
Then, the remainder is shared among all stakers -- both validator and nominators -- proportional to their stake. 
Thus, the validator receives two separate rewards: a fee for running a node, and a payout for staking. 
We remark that the commission fee is up to each validator to set, and must be publicly announced in advance. 
A higher fee translates to a higher total payout for the validator, and lower payouts to its nominators, 
so nominators will generally prefer to back validators will lower fees, and the market regulates itself in this regard. 
Validators who have built a strong reputation of reliability and performance 
will however be able to charge a higher commission fee, which is fair.

\medskip

We finalise the section with some observations on the incentives that our payout scheme is expected to cause on stakers. 
First, as validators are well remunerated and their number is limited, 
they have an incentive to ensure high backing levels from nominators to ensure getting elected, 
and thus they will value their reputation. Over time, we expect elections to be highly competitive 
and for elected validators to have strong track records of performance and reliability and large stake backings.
Second, even if payouts across different validator supports are independent of their stake, 
within a validator support each actor is paid proportional to their stake, 
so there is always an individual incentive to increase one's own stake. 
Finally, if a validator gains a particularly high level of backing, it can profit from it by either increasing 
its commission fee, which has the effect of raising its own reward at the risk of losing some nominations, 
or launching a new node as a validator candidate and splitting its backing among all its nodes. 
On this last point, we welcome operators with multiple validator nodes, 
and even aim to make their logistics simpler. 

\subsubsection{Relay-chain block limits and transaction fees}

\paragraph{Limits on resource usage:} We bound the amount of transactions that a relay-chain block can process, 
in order to a) ensure that each block can be processed efficiently even on less powerful nodes and avoid delays in block production, and b) have guaranteed availability for a certain amount of high-priority, operational transactions such as misconduct reports, even when there is high network traffic. 
In particular, we set block constraints on the following resources: on-chain byte-length, 
and time and memory required to process the transactions.

We classify transactions into several types, according to their priority level and resource consumption profile. 
For each of these types we have run tests based on worst-case scenarios for state, and for different input arguments. 
From these tests, we establish conservative estimates on resource usage for each transaction, and we use these estimates to ensure that all constraints on resource usage are observed.

We also add an extra constraint on resources: we distinguish between regular and high-priority transactions, and only let regular transactions account for up to $75\%$ of each block resource limit. This is to ensure that each block has a guaranteed space for high-priority transactions of at least $25\%$ of resources.

\paragraph{Transaction fees:} We use the model described above to set the fee level of a transaction based on three parameters: its type, its on-chain length, and its expected resource usage. This fee differentiation is used to reflect the different costs that a transaction incurs on the network and on the state, and to encourage the processing of certain types of transactions over others. A fraction of every transaction fee is paid to the block producer, while another fraction goes to finance the Treasury (Section~\ref{sec:treasury}). We highlight that, for a block producer, the rewards coming from transaction fees may constitute only a small fraction of their overall revenue, just enough to incentivise inclusion on the block.

We also run an adaptive transaction fee schedule that reacts to the traffic level, and ensures that blocks are typically far from full, so that peaks of activity can be dealt with effectively and long inclusion times are rare. In particular, the fee of each transaction is multiplied by a parameter that evolves over time depending on the current network traffic.

We make fees evolve slowly enough, so that the fee of any transaction can be predicted accurately within a frame of an hour. In particular, we do not intend for transaction fees to be the main source of income for stakers.


\subsection{Governance}\label{sec:governance}

Polkadot uses sophisticated mechanisms for Governance which allows it to evolve gracefully over time at the ultimate behest of its assembled stakeholders. A key and unfailing rule is that all changes to the protocol must be agreed upon by stake-weighted referendum -- the majority of stake always commands the network.

In order to make any changes to the network, the idea is to bring DOT holders together and administrate a network upgrade decision with the help of the Council (see Section \ref{s:council}). No matter whether the proposal is submitted by a DOT holder or by the Council, it will ultimately have to go through a referendum to let all DOT holders, weighted by stake, make the decision.

Each DOT holder in Polkadot has the right to: a) submit a proposal, b) endorse a public proposal to prioritise it in the referendum timetable, c) vote on all active referenda, d) become a candidate for a seat in the Council, and e) vote on candidates for the Council. In addition, any DOT holder may become a nominator or a validator candidate to participate in NPoS (see Section \ref{sec:validators}).

\subsubsection{Proposals and Referenda}

The core of the Polkadot logic is stored on-chain in an amorphous state-transition function and defined in a platform-neutral language: WebAssembly. Each proposal takes the form of a privileged function call in the runtime, that is able to modify the runtime code itself, achieving what would otherwise require a "hard fork". A proposal is then tabled and voted upon via referendum. 

Proposals can be started in one of several ways:
\begin{itemize}
	\item a public proposal, which is submitted by any DOT holder;
	\item a Council proposal, submitted by the Council;
	\item a proposal submitted automatically as part of the enactment of a prior referendum, and
	\item an emergency proposal submitted by the Technical Committee (Section~\ref{s:council}).
\end{itemize} 

Each proposal approved by referendum has an associated enactment delay, i.e.~a time interval between the referendum ending and the changes being enacted. For the first two types of proposals above this is a fixed interval, tentatively set to 28 days. For the third type, it can be set as desired. Emergency proposals deal with major problems with the network which need to be fast-tracked, and hence will have a shorter enactment delay. Having an enactment delay ensures a level of stability, as it gives all parties sufficient notice to adapt to the new changes. After this period, the call to the associated privileged function is automatically made.  

Any stakeholder can submit a \emph{public proposal} by depositing a fixed minimum amount of DOTs, which stays locked for a certain period. If someone agrees with the proposal, they may deposit the same amount of tokens to endorse it. Public proposals are stored in a priority queue, and at regular intervals the proposal with the most endorsements gets tabled for a referendum. The locked tokens are released once the proposal is tabled.

\emph{Council proposals} are submitted by the Council, and are stored in a separate priority queue where the priorities are set at the Council's discretion.


A \textbf{referendum} is a simple, inclusive, staked-weighted voting scheme. It has a fixed voting period, after which votes are tallied. Referenda are always binary: voting options are "aye", "nay", or abstaining entirely.

\paragraph{Timetables:} Every thirty days, a new proposal will be tabled and a referendum will come up for a vote. The proposal to be tabled is the top proposal from either the public-proposal queue or the Council-proposal queue, alternating between the two queues if both are non-empty. If both queues are empty, the slot is skipped in the referendum timetable. Multiple referenda cannot be active simultaneously, except for emergency referenda which follow a parallel timetable.

\paragraph{Vote counting:} Voting on referenda is open to all DOT holders with a voting power proportional to their stake, up to a possible vote multiplier which is awarded to some parties depending on their level of commitment to the system, as we explain now. A party must generally lock their tokens used for voting until at least the enactment delay period beyond the end of the referendum. This is in order to ensure that some minimal economic buy-in to the result is needed and to dissuade vote selling. It is possible to vote without locking at all, but in that case the voting power is a small fraction of a normal vote for the given stake. Conversely, Polkadot will offer \emph{voluntary extended locking}, that allows any party to increase their voting power by extending the period of time they are willing to lock up their tokens. This ensures that voters committed to the system long term, who are willing to increase their exposure to the decision of a referendum, have a greater say in the matter. 

\paragraph{Turnout biasing:} It may seem restrictive to force a full stakeholder-based process to do something as little as, say, nudging the block time down by $5\%$. However, without this rule the network would likely be unstable, as placing its control outside of the hands of stakeholders would create a misalignment that may lead to inaction or worse. However, by taking advantage of the fact that turnout is rarely $100\%$, we can effect different outcomes depending on the circumstances, crafting a balance of power between active and passive stakeholders. For example, simple voting systems typically introduce a notion of quorum, whereby a minimum amount of turnout must be reached before a change is passed. 

For public proposals, we generalise this notion into a "positive turnout bias", where additional turnout always makes change more likely, assuming the same yay-to-nay ratio. More specifically, in case of low turnout we favour the nay side, or status quo, by requiring a super-majority approval, and as turnout approaches $100\%$ the requirement dials down to majority-carries. This works on two principles: Firstly that the status quo tends to be safer than any change, and thus should have some bias towards it. Secondly that, like all means of empirical measurement, there is inevitably going to be some degree of inaccuracy and volatility over time, particularly when turnout is low -- a result could be $51\%-49\%$ one month and then change to $49\%-51\%$, and given the costs involved in enacting the changes of a proposal it is advantageous to ensure that a result would not likely flip shortly after enactment. 

On the other hand, for proposals submitted by the Council, referenda have no turnout bias and majority-carries is observed. The reasoning here is that proposals pre-approved by the Council are deemed safer and less likely to be reverted, so the previously mentioned issues are alleviated and we can let DOT holders freely decide on the matter. 

\begin{figure}[htb]
	\centering
	\includegraphics[width=1.1\textwidth]{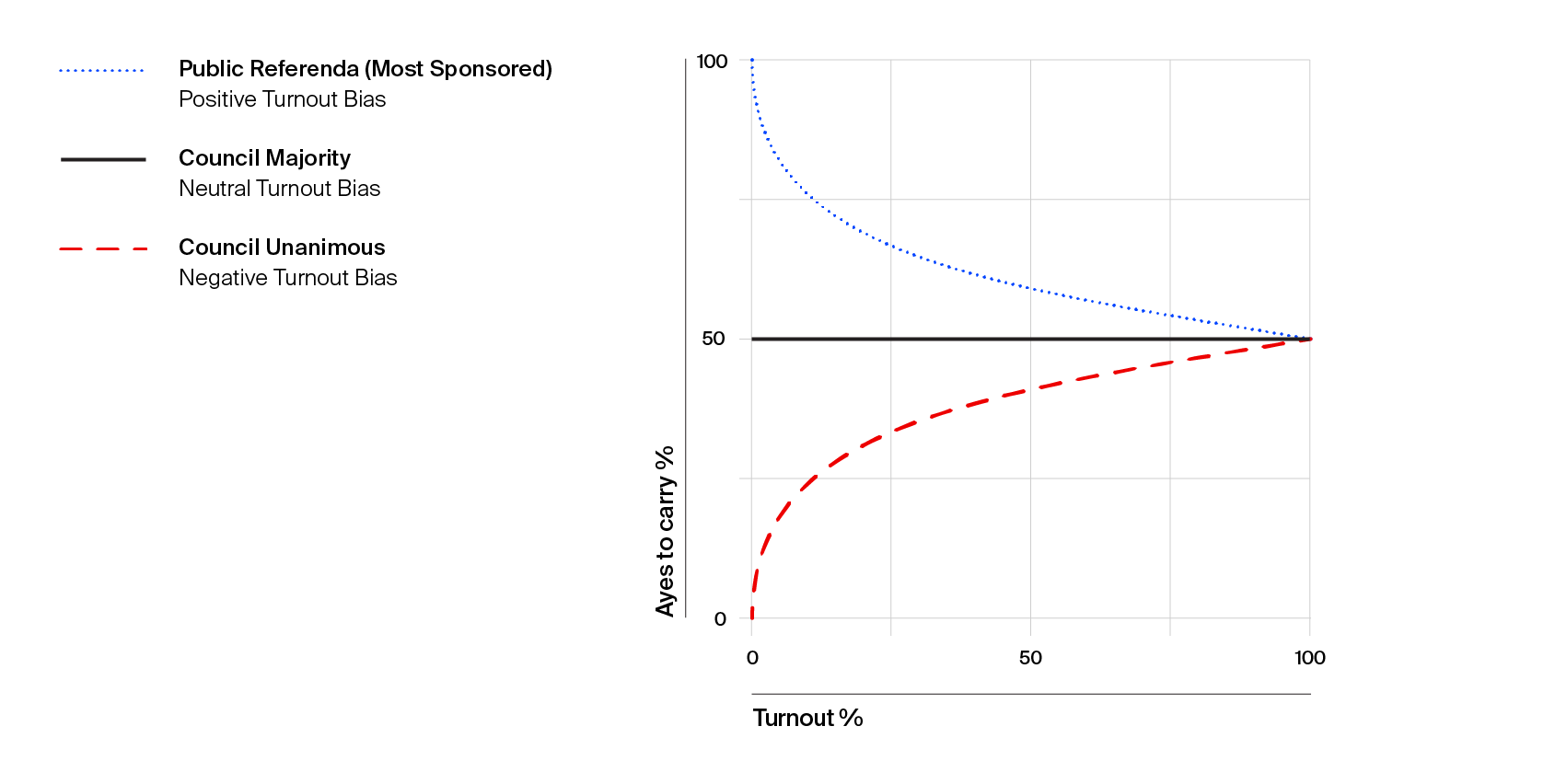}
	\caption{Adaptive turnout biasing (image credit: Ignasi Albero).}
	\label{fig:biasing}
\end{figure}


Finally, in the exceptional case that a Council proposal receives unanimous support by all Council members, it will observe a "negative turnout bias". This is the symmetric opposite of the first case, where additional turnout always makes change less likely, we favour the yay side in case of low turnout by requiring a super-majority of nays to reject the proposal, and as turnout approaches $100\%$ the requirement dials down to majority-carries. See Figure~\ref{fig:biasing}.

\subsubsection{The Council and the Technical Committee}\label{s:council}

\textbf{The Council} is an entity comprising a number of actors each represented by an on-chain account. Its goals are to represent passive stakeholders, submit sensible and important proposals, and cancel uncontroversially dangerous or malicious proposals.

The Council will constantly review candidate proposals to deal with emerging issues in the system. A candidate proposal is officially backed by the Council -- and enters the queue of Council proposals -- only after it is approved by a strict majority of Council members, with no member exercising a veto. A candidate proposal can be vetoed only once; if, after a cool-down period, it is once more approved by a majority of Council members, it cannot be vetoed a second time. 


As mentioned before, in the case that all members vote in favour, a Council proposal is considered uncontroversial and enjoys a special turnout bias that makes it more likely to be approved. 

Finally, Council members may vote to cancel any proposal, regardless of who submitted it, but their vote must be unanimous. Since unanimity is a high requirement, it is expected that this measure will only be used when it is an entirely uncontroversial move. This may function as a last resort if there is an issue found late in the day with a referendum's proposal such as a bug in the code of the runtime or a vulnerability  that the proposal would institute. If the cancellation is controversial enough that there is at least one dissenter, then it will be left to all DOT holders en masse to determine the fate of the proposal, with the registered Council cancellation votes serving as red flags so that voters pay special attention.

\paragraph{Electing Council members:} There will be 23 seats in the Council of Polkadot. A general election for all seats will take place once per month. 
All DOT holders are free to register their candidacy for the Council, and free to vote for any number of candidates, with a voting power proportional to their stake. Much like the validator election problem in NPoS, this is a stake-weighted, multi-winner election problem based on approval ballots. We can thus solve it using the same algorithm we use for NPoS, which in particular offers the property of \emph{proportional justified representation}; see Section~\ref{sec:validators} for more details. This property guarantees that the elected Council will represent as many minorities as possible, thus ensuring that Governance stay decentralised and resistant to capture. Council members can be re-elected indefinitely, provided their approval remains high enough. 

\medskip

\textbf{The Technical Committee} is composed according to a single vote for each team that has successfully and independently implemented or formally specified the protocol in Polkadot, or in its canary network Kusama\footnote{http://kusama.network}. Teams may be added or removed by a simple majority of the Council. 

The Technical Committee is the last line of defence for the system. Its sole purpose is detecting present or imminent issues in the system such as bugs in the code or security vulnerabilities, and proposing and fast-tracking emergency referenda. An emergency proposal needs a simultaneous approval of at least three-quarters of the Council members and at least two-thirds of the Technical Committee members in order to be tabled. Once tabled, it is fast-tracked into a referendum that runs in parallel with the timetable of regular referenda, with a far shorter voting period, and a near-zero enactment period. The approval mechanics in this case are unchanged from what they would be otherwise, i.e.~either a simple majority or, in the case of a unanimous Council approval, a turnout-based bias for approval.

We highlight that for practical reasons the Technical Committee is not democratically elected, but in contrast it has an extremely reduced scope of action and no power to act unilaterally, as explained in the lines above. This mechanism is expected to suffice for non-contentious bug fixes and technical upgrades, but given the requirements imposed, may not be effective in the case of emergencies that have a tinge of political sensitivity or strategic importance to them. 

\subsubsection{Allocation of parachain slots}\label{s:pAllocation}

We use auctions to have a fair, transparent and permissionless parachain allocation procedure. 
Broadly speaking, parties interested in receiving a parachain slot participate in an auction with DOT-denominated bids. The party with the highest bid is declared as winner and is allocated a slot for a specific period of time, with its bid becoming a locked deposit that is released at the end of said period. The leasing cost of the slot thus corresponds to the opportunity cost of having this deposit locked. This DOT-denominated deposit also establishes the voting power of the parachain in Polkadot's governance.

Since implementing seal-bid auctions is difficult and in order to avoid bid sniping, we adopt a Candle auction \cite{Fuellbrunn:2012:CandleAuction} mechanism with a retroactively determined close. 
Going into detail, we plan to have an auction every few weeks, where in each auction four contiguous six-month slots are offered for lease. A bid can be made for any combination of one, two, three or four contiguous slots, for a total of ten possible time periods lasting 6, 12, 18 or 24 months. Once the auction starts, parties can post bids as transactions for any one of these ten periods, within a fixed window of time lasting several hours. A party is allowed to submit multiple bids, where a bid is registered only if a) it beats the current highest bid for the corresponding period, and b) the party does not become the provisional winner of two or more periods with gaps in between. For example, the winner of period $(1,2)$ -- constituted of the first two slots -- cannot bid on period $(4)$ -- the fourth slot -- until someone else overbids the former period. 


The stated goals of this design are to incentivise parties to bid early and avoid bid sniping, to give less funded projects a chance of winning a slot hence securing the decentralised nature of Polkadot, and to discourage griefing attacks by parties who raise the value of the winning bid with no intention of winning themselves.

\subsubsection{Treasury}\label{sec:treasury}

The Treasury raises funds continually. These funds are used to pay for developers that provide software updates, apply any changes decided by referenda, adjust parameters, and generally keep the system running smoothly. Funds may also be used for further goals such as marketing activities, community events and outreach. This is ultimately controlled by all DOT holders via Governance and it will be the community and their collective imagination and judgment which really determines the course of the Treasury.

Funds for Treasury are raised in two ways:

\begin{enumerate}
	\item by channeling some of the validator rewards that come from minting of new tokens, and 
	\item by channeling a fraction of transaction fees and of slashings.
\end{enumerate}

The first method allows us to maintain a fixed inflation rate while simultaneously having the validator rewards be dependent of the staking level (see Section~\ref{sec:inflation}): the difference between the scheduled minted tokens and the validator rewards is assigned to Treasury in each era. 
We also argue that it is convenient to have a fraction of all slashings be redirected to Treasury: following an event that produced heavy stake slashing, the system is likely to need additional funds to develop software updates or new infrastructure that deal with an existing issue, or it might be decided by Governance to reimburse some of the slashed stake. Thus, it makes sense to have the slashed DOTs available in Treasury, instead of burning them and having to mint more DOTs soon thereafter.

\subsection{Cryptography}\label{sec:crypto}

In Polkadot, we necessarily distinguish among different permissions and functionalities with different keys and key types, respectively.  We roughly categorise these into account keys with which users interact and session keys that nodes manage without operator intervention beyond a certification process.

\subsubsection{Account keys}

Account keys have an associated balance of which portions can be {\em locked} to play roles in staking, resource rental, and governance, including waiting out a couple types of unlocking period.  We allow several locks of varying duration, both because these roles impose different restrictions, and for multiple unlocking periods running concurrently. 

We encourage active participation in all these roles, but they all require occasional signatures from accounts.  At the same time, account keys have better physical security when kept in inconvenient locations, like safety deposit boxes, which makes signing arduous.  We avoid this friction for users as follows.

Accounts that lock funds for staking are called {\em stash accounts}.  All stash accounts register a certificate on-chain that delegates all validator operation and nomination powers to some {\em controller account}, and also designates some {\em proxy key} for governance votes.  In this state, the controller and proxy accounts can sign for the stash account in staking and governance functions respectively, but not transfer funds.  

\smallskip

At present, we support both ed25519 \cite{ed25519} and Schnorrkel/sr25519 \cite{schnorrkel} for account keys.  These are both Schnorr-like signatures implemented using the Ed25519 curve, so both offer extremely similar security.  We recommend ed25519 keys for users who require Hardware Security Module (HSM) support or other external key management solution, while Schnorrkel/sr25519 provides more blockchain-friendly functionality like Hierarchical Deterministic Key Derivation (HDKD) and multi-signatures.  

In particular, Schnorrkel/sr25519 uses the Ristretto implementation \cite{Ristretto} of Mike Hamburg's Decaf \cite[\S7]{Decaf}, which provide the 2-torsion free points of the Ed25519 curve as a prime order group.  Avoiding the cofactor like this means Ristretto makes implementing more complex protocols significantly safer.  We employ Blake2b for most conventional hashing in Polkadot, but Schnorrkel/sr25519 itself uses STROBE128 \cite{STROBE}, which is based on Keccak-f(1600) and provides a hashing interface well suited to signatures and non-interactive zero-knowledge proofs (NIZKs).

\subsubsection{Session keys}\label{sec:session_keys}

Session keys each fill roughly one particular role in consensus or security.  As a rule, session keys gain authority only from a session certificate, signed by some controller key, that delegates appropriate stake.  

At any time, the controller key can pause or revoke this session certificate and/or issue replacement with new session keys.  All new session keys can be registered in advance, and most must be, so validators can cleanly transition to new hardware by issuing session certificates that only become valid after some future session.  We suggest using pause mechanism for emergency maintenance and using revocation if a session key might be compromised.  

We prefer if session keys remain tied to one physical machine because doing so minimises the risk of accidental equivocation.  We ask validator operators to issue session certificates using an RPC protocol, not to handle the session secret keys themselves.  

Almost all early proof-of-stake networks have a negligent public key infrastructure that encourages duplicating session secret keys across machines, and thus reduces security and leads to pointless slashing.

\smallskip

We impose no prior restrictions on the cryptography employed by specific components or their associated session keys types.\footnote{We always implement cryptography for Polkadot in native code, not just because the runtime suffers from WASM's performance penalties, but because all of Polkadot's consensus protocols are partially implemented outside the runtime in Substrate modules.}

In BABE \ref{sec:babe}, validators use Schnorrkel/sr25519 keys both for regular Schnorr signatures, as well as for a verifiable random function (VRF) based on NSEC5 \cite{NSEC5}.  

A VRF is the public-key analog of a pseudo-random function (PRF), aka cryptographic hash function with a distinguished key, such as many MACs.  We award block production slots when the block producer scores a low enough VRF output $\mathtt{VRF}_{\sk}(r_e || \mathtt{slot\_number} )$, so anyone with the VRF public keys can verify that blocks were produced in the correct slot, but only the block producers know their slots in advance via their VRF secret key.

As in \cite{Praos}, we provide a source of randomness $r_e$ for the VRF inputs by hashing together all VRF outputs form the previous session, which requires that BABE keys be registered at least two full epochs before being used.

We reduce VRF output malleability by hashing the signer's public key alongside the input, which dramatically improves security when used with HDKD.  We also hash the VRF input and output together when providing output used elsewhere, which improves composability when used as a random oracle in security proofs.  See the 2Hash-DH construction from Theorem 2 on page 32 in appendix C of \cite{Praos}.  

In GRANDPA \ref{sec:grandpa}, validators shall vote using BLS signatures, which supports convenient signature aggregation and select ZCash's BLS12-381 curve for performance.  There is a risk that BLS12-381 might drop significantly below 128 bits of security, due to number field sieve advancements.  If and when this happens, we expect upgrading GRANDPA to another curve to be straightforward. 


We treat libp2p's transport keys roughly like session keys too, but they include the transport keys for sentry nodes, not just for the validator itself.  As such, the operator interacts slightly more with these.


\subsection{Networking}\label{sec:networking}

In the preceding sections we talk about nodes sending data to another node or
other set of nodes, without being too specific on how this is achieved. We do
this to simplify the model and to clearly delineate a separation of concerns
between different layers.

Of course, in a real-world decentralised system the networking part also must
be decentralised - it's no good if all communication passes through a few
central servers, even if the high-level protocol running on top of it is
decentralised with respect to its entities. As a concrete example: in certain
security models, including the traditional Byzantine fault-tolerant setting,
nodes are modelled as possibly malicious but no consideration is given to
malicious edges. A security requirement like “$> 1/3$ of nodes are honest” in
the model, in fact translates to “$> 1/3$ of nodes are honest and can all
communicate perfectly reliably with each other all the time” in reality.
Conversely, if an edge is controlled by a malicious ISP in reality, it is the
corresponding node(s) that must be treated as malicious in any analysis under
the model. More significantly, if the underlying communications network is
centralised, this can give the central parties the ability to “corrupt” $> 1/3$
of nodes within the model thereby breaking its security assumptions, even if
they don't actually have arbitrary execution rights on that many nodes.

In this section we outline and enumerate the communication primitives that we
require in Polkadot, and sketch a high-level design on how we achieve these in
a decentralised way, with the specifics to be refined as we move forward with a
production system.

\subsubsection{Networking overview}

As discussed above, Polkadot consists of a unique relay chain interacting with
many different parachains and providing them with security services. These
require the following network-level functionality, generally for distribution
and availability:

\begin{enumerate}
	
	\item As with all blockchain-like protocols, the relay chain requires:
	\begin{enumerate}
		\item accepting transactions from users and other external data (collectively known as extrinsic data or \emph{extrinsics}), and distributing them
		\item distributing artefacts of the collation subprotocol \ref{sec:relaychainblockproduction}
		\item distributing artefacts of the finalisation subprotocol \ref{sec:grandpa}
		\item synchronising previously-finalised state
	\end{enumerate}
	
	As an important special case, parachains may choose to implement themselves
	according to the above structure, perhaps even re-using the same subprotocols.
	Part of the Polkadot implementation is architected as a separate library called
	\texttt{substrate} for them to do just this.
	
	\item For interaction between the relay chain and the parachains, we require:
	\begin{enumerate}
		\item accepting parachain blocks from parachain collators
		\item distributing parachain block metadata including validity attestations
		\item distributing parachain block data and making these available for a time \ref{sec:validity-and-availability}, for auditing purposes
	\end{enumerate}
	
	\item For interaction between parachains, we require:
	\begin{enumerate}
		\item distributing messages between parachains \ref{sec:XCMP}, specifically from the relevant senders to the relevant recipients
	\end{enumerate}
\end{enumerate}

For each of the above functionality requirements, we satisfy them with the
following:

\begin{itemize}
	\item 1(b), 1(c), 2(b) - artefacts are broadcast as-is (i.e. without further coding) via \hyperref[sec:gossiping]{gossip}.
	\item 1(a), 1(d), 2(a) - effectively, a set of nodes provide the same \hyperref[sec:net_service]{distributed service} to clients. For accepting extrinsics or blocks, clients send these directly to the serving node; for synchronisation, clients receive verifiable data directly from the serving node.
	\item 2(c) - special case, \hyperref[sec:net_storage]{below}. Briefly, data is erasure-encoded so that different recipients receive a small part; pieces are sent directly via QUIC.
	\item 3(a) - special case, \hyperref[sec:net_crosschain]{below}. Briefly, messages are sent directly via QUIC; in the process outboxes are reassembled into inboxes and the latter can be transferred as a batch to recipients.
\end{itemize}

We go into these in more detail in the next few sections. Finally, we talk about the lower layers underpinning all of these subprotocols, namely \nameref{sec:net_lowlevel}.

\subsubsection{Gossiping} \label{sec:gossiping}

This subprotocol is used for most relay-chain artefacts, where everyone needs to see more-or-less the same public information. Part of its structure is also used for when a node goes offline for a long time and needs to synchronise any newer data it hasn't seen before.

The Polkadot relay chain network forms a gossip overlay network on top of the physical communications network, as an efficient way to provide a decentralised broadcast medium. The network consists of a known number of trusted nodes (validators) who have been permissioned via staking, and an unknown number of untrusted nodes (full nodes that don't perform validation) from the permissionless open internet. (As an aside, recall that some of the untrusted nodes may have other roles as defined earlier, e.g. parachain collator, fishermen, etc.)

A simple push-based approach is implemented currently, with hash-based tracker caches to avoid sending duplicates to peers, and a few restrictions to avoid the most common spam attacks:

\begin{itemize}
	\item Artefacts may only be received in dependency order; peers are not allowed to send them out-of-order. Though this decreases network-level efficiency, it is straightforward to implement and provides a healthy level of security.
	\item To efficiently communicate to sending peers what they are allowed to send in dependency order, periodically peers update each other with their view of the latest heads of the chain.
\end{itemize}

There are also more specific constraint rules applied to artefacts belonging to the various higher-level subprotocols using the gossip protocol, to avoid broadcasting obsolete or otherwise unneeded artefacts. For example, for GRANDPA we only allow two votes being received for each type of vote, round number, and voter; any further votes will be ignored. For block production only valid block producers are allowed to produce one block per round; any further blocks will be ignored.

There is basic support for \emph{sentry nodes}, proxy servers that are essentially the only neighbour of a private server, running more security-critical operations such the validator role.

The network topology is a weak point currently; nodes connect to each other on an ad-hoc basis by performing random lookups in the \hyperref[sec:net_lowlevel]{address book}. Further work will proceed along two fronts:

\begin{enumerate}
	\item Trusted nodes will reserve a portion of their bandwidth and connection resources, to form a structured overlay with a deterministic but unpredictable topology that rotates every era. For nodes running behind sentries, this effectively means that their sentry nodes instead participate in this topology.
	
	\item For the remainder of trusted nodes' resource capacity, and for the whole of untrusted nodes' resource capacity, they will select neighbours via a scheme based on latency measurements, with the details to be decided. Notably, for good security properties we want a scheme that does not simply choose "closest first", but also some far links as well.
\end{enumerate}

In some sense, this can be viewed as the trusted nodes forming a core with the untrusted nodes around it - but note that trusted nodes are expected to use some of their resources to serve untrusted nodes as well. Both topologies are chosen to mitigate eclipse attacks, as well as sybil attacks in the permissionless untrusted case.

Further work will also include some sort of set reconciliation protocol, to further reduce redundancy when many senders attempt to send the same object to the same recipient at once; and potentially look into lifting the dependency-order restriction whilst retaining security.

\subsubsection{Distributed service} \label{sec:net_service}

This subprotocol is used when some part of Polkadot is providing a service to some external entity, namely 1(a) accepting relay chain transactions, 1(d) synchronising relay chain state, and 2(a) accepting collated blocks in the list above.

In our initial implementation, this simply involves looking up a particular target set in the address book, selecting a few nodes from this set, and connecting to them. For 1(a) and 1(d) the target set is the whole set of validators, and for 2(a) the target set is the set of parachain validators for the client collator's parachain. Both of these can be retrieved straightforwardly from the chain state, and indeed for 1(a) this is simply the same process as joining the gossip network.

Further work will consider the issue of load-balancing the transport-layer connections over the whole target set, as well as ensuring availability. This may require additional sophistication in the address book.

\subsubsection{Storage and availability} \label{sec:net_storage}

This subprotocol addresses the networking used for the Availability and Validity subprotocol described in Section \ref{sec:validity-and-availability}.

Recall that for scalability, Polkadot does not require everyone to store the state of the whole system, namely all of the state pointed to by all of the blocks. Instead, every parachain block is split into pieces by erasure-coding, such that there is 1 piece for every validator for a total of $N$ pieces, the erasure threshold being $ceil(N/3)$ blocks for security reasons explained elsewhere. All the pieces are available initially at the relevant parachain validators, having been submitted by some of the collators. (In this role, the parachain validators are also called the \emph{preliminary checkers}.) The pieces are then selectively distributed in the following phases:

\begin{enumerate}
	\item Distribution - every validator initially wants 1 of these pieces, and the parachain validators must distribute them as such.
	\item Retrieval - the \emph{approval checkers} need to be convinced that $ceil(N/3)$ validators have their pieces, and some of them will attempt to actually retrieve these.
	\item Further retrieval - yet later, and optionally, other non-validator parties might also want to perform further checks, e.g. in response to fishermen alerts, and again will want any $ceil(N/3)$ of the pieces.
\end{enumerate}

This subprotocol therefore aims to ensure that the threshold is available and can be retrieved from the relevant validators for some reasonable amount of time, until at least the latter phases are complete. We will follow a bittorrent-like protocol with the following differences:

\begin{itemize}
	\item With both distribution and retrieval, the set of recipients is known. Therefore, pieces can be pre-emptively pushed from validators that already have the piece, in addition to bittorrent's pull semantics.
	\item Validators behind sentry nodes will use these as proxies, rather than directly sending.
	\item Instead of a centralised tracker, tracker-like information such as who has what piece, is broadcast via the relay chain \hyperref[sec:gossiping]{gossip network}.
\end{itemize}

The preliminary checkers are expected to be fully- or mostly-connected; this is a pre-existing requirement for the collation protocol as well. The approval checkers should also be fully- or mostly-connected, to help the retrieval process complete faster.

Beyond this, nodes may communicate to any other nodes as the protocol sees fit, similar to bittorrent. To protect against DoS attacks they should implement resource constraints as in bittorrent, and furthermore nodes should authenticate each other and only communicate with other validators, including the preliminary and approval checkers. Non-validator parties in the latter optional phase will be supplied with an authentication token for this purpose. In a separate more detailed document, we propose a scheme for load-balancing, such that nodes do not accidentally overwhelm other nodes in their random choices.

There are various reasons why we do not consider it very suitable, to use a
structured overlay topology for this component:

\begin{enumerate}
	\item Each piece is sent to specific people, rather than everyone.
	\item
	\begin{enumerate}
		\item People that want a specific piece of data, know where to get it - i.e.
		validators, for their own piece, i.e. the preliminary checkers.
		\item Other people want non-specific pieces - i.e. approval checkers,
		want any 1/3 of all pieces to be able to reconstruct.
	\end{enumerate}
\end{enumerate}

Overlay topologies are generally more useful for the exact opposite of the
above usage requirements:

\begin{enumerate}
	\item Each data piece is sent to nearly everyone, or
	\item People want a specific data piece, but don't know where to get it from.
\end{enumerate}

For example, bittorrent has similar requirements and does not use a structured
overlay - the peers there connect to other peers on a by-need basis.

\subsubsection{Cross-chain message} \label{sec:net_crosschain}

This subprotocol address the networking scheme used for the XCMP messaging sub-protocol described in Section \ref{sec:XCMP}.

To recap that section, parachains can send messages to each other. The contents
of outgoing messages are part of the parachain PoV blocks submitted by the
sending parachain collators to its validators. This is distributed to other
validators as part of the \hyperref[sec:net_storage]{availability protocol}.
Relay chain blocks contain metadata that describes the relevant outgoing
messages corresponding to the incoming messages for every parachain. The job of
XCMP networking therefore, is for each recipient parachain to obtain its
incoming messages from the outboxes.

Note that without any additional sophistication, this can always be done be
retrieving the erasure-coded pieces of the A\&V protocol, for example via the
\hyperref[sec:gossiping]{gossip network}, and decoding all the outboxes of all
potential senders. This of course is very inefficient - both using a broadcast
medium for data only the recipient parachain is interested in, and in the data
being retrieved which includes messages sent to parachains other than the
recipient. Nevertheless this can serve as an initial naive implementation for
the early stages of the network where traffic is expected to be low.

Further work will proceed along the following lines. One view of the problem is
how to efficiently convert the outboxes of all senders, into the inboxes of all
recipients. Once we have done that, any recipient can simply retrieve the inbox
directly, from whomever has done the conversion. We note that the structure of
our A\&V networking has a very similar communication requirement - there, the
pieces of each parachain block have to be distributed to every other validator,
and conversely every validator has to receive a piece of every parachain block.
Therefore, our main efforts will be directed towards extending the A\&V
networking protocol, to support the conversion of XCMP outboxes into inboxes.

One important difference that we must cover, is that the pieces in A\&V have
some in-built redundancy, whereas XCMP messages have no in-built redundancy and
must all be distributed reliably. Applying erasure coding to these as well, is
a straightforward and obvious solution, but we will also explore alternatives.

\subsubsection{Sentry nodes} \label{sec:net_sentry}

Sometimes, network operators want to arrange aspects of their physical network
for operational security reasons. Some of these arrangements are independent
and compatible with the design of any decentralised protocol, which typically
works in the layer above. However some other arrangements need special
treatment by the decentralised protocol, in particular arrangements affecting
the reachability of nodes.

For such use-cases, Polkadot supports running full-nodes as the sentry nodes of
another full-node that is only reachable by these sentry nodes. This works best
when one runs several sentry nodes for a single private full-node. Protocol
wise, briefly, sentry nodes are regular neighbours of their private node, with
some additional metadata to tell others to communicate this private node via
its sentry nodes. In direct-sending mode, they act similarly to TURN servers,
without any resource bottleneck constraints since every sentry node is serving
only one private node. These additions are fairly straightforward and more
details are available elsewhere.

It is not required to run sentry nodes, for example if you believe the
aforementioned security benefits are not worth the added latency cost.

A brief discussion about the security tradeoffs of this approach follows. One
benefit of a restricted physical topology, is to support load-balancing and DoS
protection across a number of gateway proxies. The indirection can also help to
protect the private node if there are exploits in the software - although note
this does not cover the most severe exploits that give arbitrary execution
rights on the sentry node, which can then be used as a launching pad for
attacks on the private node. So, while we don't believe that a public address
is itself a security liability when the serving code is written well, sentry
nodes can help to mitigate these other scenarios.

(An alternative possibility is for the network operator to run lower-level
proxies, such as IP or TCP proxies, for their private node. This certainly can
be done without any protocol support from Polkadot. One advantage of sentry
nodes compared to this scenario is that the traffic coming from a sentry node
has been through some level of verification and sanitisation as part of the
Polkadot protocol, which would be missing for a lower-level proxy. Of course
there can be exploits against this, but these are dealt with with high priority
since they are amplification vectors usable against the whole network.)

\subsubsection{Authentication, transport, and discovery} \label{sec:net_lowlevel}

In secure protocols in general, and likewise with Polkadot, entities refer to each other by their cryptographic public keys. There is no strong security association with weak references such as IP addresses since those are typically controlled not by the entities themselves but by their communications provider.

Nevertheless in order to communicate we need some sort of association between entities and their addresses. In Polkadot we use a similar scheme as many other blockchains do, that is using the widely used distributed hash table (DHT), Kademlia \cite{Maymounkov:2002:Kademila}. Kademlia is a DHT that uses the XOR distance metric, and is often used for networks with high churn. We use Protocol Labs' libp2p Kademlia implementation with some changes for this purpose. To prevent Eclipse attacks \cite{eclipseattack} we allow for routing tables that are large enough to contain at least some honest nodes, and are in the process of implementing the S-Kademia approach for multipath routing.

Currently, this \emph{address book} service is also used as the primary discovery mechanism - nodes perform random lookups on the space of keys when they join the network, and connect to whichever set of addresses are returned. Likewise, nodes accept any incoming connections. This makes it easy to support light clients and other unprivileged users, but also makes it easy to perform DoS attacks.

Further work will decouple the discovery mechanism from the address book, as described in \nameref{sec:gossiping}, resulting in a more security network topology. Part of this will require some fraction of transport-level connections be authenticated against the currently-trusted validator set. However we also require to retain the ability to accept incoming connections from unauthenticated entities, and this needs to be restricted on a resource basis, without starving the authenticated entities.

Further work will also decouple the implementation of the address book from its interface, so that e.g. we can put part of in on-chain. This has different security tradeoffs from a Kademlia-based address book, some of which is outside of the current scope of Polkadot, such as location privacy. By offering different possibilities, we hope to satisfy a diverse set of nodes having different security requirements.


\section{Future Work}\label{sec:futurework}
For future work we plan to focus on a number of extensions for Polkadot.
We want to add an incentivisation model for fishermen to make sure they are incentivised to report malicious behaviour. 
To enable trustless messaging we are working on SPREE \ref{sec:SPREE}. We are also interested to increase scalability of Polkadot further, for example by investigating the idea of having nested relay chains. Moreover, we are working on incorporating bridging protocols \ref{sec:bridge} to other chains  such as Bitcoin, Ethereum, and Zcash. Furthermore, to increase usability we are planning to enable \emph{Parathreads}, which have the same utility as parachains but are temporary and have a different fee model.

.




\section*{Acknowledgement}
We would like to thank Bill Laboon from Web3 Foundation for his feedback and Parity Technologies developers for their useful input and good discussions.
\bibliographystyle{plain}
\bibliography{references,crypto,rt,grandpa}

\begin{thebibliography}{10}

\bibitem{availandvalid}
Availability and validity scheme.
\newblock
  \url{https://research.web3.foundation/en/latest/polkadot/Availability_and_Validity/}.

\bibitem{babe}
Blind assignment for blockchain extension (babe).
\newblock \url{https://research.web3.foundation/en/latest/polkadot/BABE/Babe/}.

\bibitem{Visa2020}
Visa inc. at a glance, Dec 2015.
\newblock Accessed: 2020-02-25.

\bibitem{availabilityETH2}
Mustafa Al-Bassam, Alberto Sonnino, and Vitalik Buterin.
\newblock Fraud and data availability proofs: Maximising light client security
  and scaling blockchains with dishonest majorities.
\newblock {\em arXiv preprint arXiv:1809.09044}, 2018.

\bibitem{consensusonclock}
Handan~K{\i}l{\i}n\c{c} Alper.
\newblock Consensus on clock in universally composable timing model.
\newblock Cryptology ePrint Archive, Report 2019/1348, 2019.
\newblock \url{https://eprint.iacr.org/2019/1348}.

\bibitem{ed25519}
Daniel Bernstein, Niels Duif, Tanja Lange, Peter Schwabe, and Bo-Yin Yang.
\newblock High-speed high-security signatures.
\newblock {\em Journal of Cryptographic Engineering volume}, 2012(2):77--89,
  2012.

\bibitem{brill2017phragmen}
Markus Brill, Rupert Freeman, Svante Janson, and Martin Lackner.
\newblock Phragm{\'e}n’s voting methods and justified representation.
\newblock In {\em Thirty-First AAAI Conference on Artificial Intelligence},
  2017.

\bibitem{Tendermint}
Ethan Buchman, Jae Kwon, and Zarko Milosevic.
\newblock The latest gossip on bft consensus.
\newblock {\em arXiv preprint arXiv:1807.04938}, 2018.

\bibitem{schnorrkel}
Jeffrey Burdges.
\newblock schnorrkel: Schnorr vrfs and signatures on the ristretto group.
\newblock \url{https://github.com/w3f/schnorrkel}, 2019.

\bibitem{buterin2014ethereum}
Vitalik Buterin.
\newblock Ethereum: A next-generation smart contract and decentralized
  application platform, 2014.
\newblock Accessed: 2016-08-22.

\bibitem{CasperFFG}
Vitalik Buterin and Virgil Griffith.
\newblock Casper the friendly finality gadget.
\newblock {\em arXiv preprint arXiv:1710.09437}, 2017.

\bibitem{NPoSpaper}
Alfonso Cevallos and Alistair Stewart.
\newblock Validator election in nominated proof-of-stake.
\newblock {\em arXiv preprint arXiv:2004.12990}, 2020.

\bibitem{chitra2019competitive}
Tarun Chitra.
\newblock Competitive equilibria between staking and on-chain lending.
\newblock {\em arXiv preprint arXiv:2001.00919}, 2019.

\bibitem{scaling}
Kyle Croman, Christian Decker, Ittay Eyal, Adem~Efe Gencer, Ari Juels, Ahmed
  Kosba, Andrew Miller, Prateek Saxena, Elaine Shi, Emin Sirer, Dawn Song, and
  Roger Wattenhofer.
\newblock On scaling decentralized blockchains.
\newblock volume 9604, pages 106--125, 02 2016.

\bibitem{Praos}
Bernardo David, Peter Ga{\v{z}}i, Aggelos Kiayias, and Alexander Russell.
\newblock Ouroboros {P}raos: An adaptively-secure, semi-synchronous
  proof-of-stake blockchain.
\newblock In {\em Annual International Conference on the Theory and
  Applications of Cryptographic Techniques}, pages 66--98. Springer, 2018.
\newblock \url{https://eprint.iacr.org/2017/573}.

\bibitem{Fuellbrunn:2012:CandleAuction}
Sascha Füllbrunn and Abdolkarim Sadrieh.
\newblock {Sudden Termination Auctions—An Experimental Study}.
\newblock {\em Journal of Economics \& Management Strategy}, 21(2):519--540,
  June 2012.

\bibitem{backbone}
Juan Garay, Aggelos Kiayias, and Nikos Leonardos.
\newblock The bitcoin backbone protocol: Analysis and applications.
\newblock In {\em Annual International Conference on the Theory and
  Applications of Cryptographic Techniques}, pages 281--310. Springer, 2015.

\bibitem{Decaf}
Mike Hamburg.
\newblock Decaf: Eliminating cofactors through point compression.
\newblock In Rosario Gennaro and Matthew Robshaw, editors, {\em Advances in
  Cryptology -- CRYPTO 2015}, pages 705--723, Berlin, Heidelberg, 2015.
  Springer Berlin Heidelberg.
\newblock \url{https://eprint.iacr.org/2015/673}.

\bibitem{STROBE}
Mike Hamburg.
\newblock The {STROBE} protocol framework.
\newblock IACR ePrint 2017/003, 2017.
\newblock \url{https://eprint.iacr.org/2017/003} and
  \url{https://strobe.sourceforge.io}.

\bibitem{eclipseattack}
Ethan Heilman, Alison Kendler, Aviv Zohar, and Sharon Goldberg.
\newblock Eclipse attacks on bitcoin's peer-to-peer network.
\newblock In {\em 24th {USENIX} Security Symposium ({USENIX} Security 15)},
  pages 129--144, Washington, D.C., August 2015. {USENIX} Association.

\bibitem{Ristretto}
Isis Lovecruft and Henry de~Valence.
\newblock Ristretto.
\newblock \url{https://ristretto.group}.
\newblock Accessed: 2019.

\bibitem{Maymounkov:2002:Kademila}
Petar Maymounkov and David Mazi\`{e}res.
\newblock Kademlia: A peer-to-peer information system based on the xor metric.
\newblock In {\em Revised Papers from the First International Workshop on
  Peer-to-Peer Systems}, IPTPS '01, pages 53--65, London, UK, UK, 2002.
  Springer-Verlag.

\bibitem{ALGORAND}
Silvio Micali.
\newblock {ALGORAND:} the efficient and democratic ledger.
\newblock {\em CoRR}, abs/1607.01341, 2016.

\bibitem{vrf}
Silvio Micali, Michael Rabin, and Salil Vadhan.
\newblock Verifiable random functions.
\newblock In {\em 40th Annual Symposium on Foundations of Computer Science
  (Cat. No. 99CB37039)}, pages 120--130. IEEE, 1999.

\bibitem{ntp}
David Mills et~al.
\newblock Network time protocol.
\newblock Technical report, RFC 958, M/A-COM Linkabit, 1985.

\bibitem{nakamoto2008bitcoin}
Satoshi Nakamoto.
\newblock Bitcoin: A peer-to-peer electronic cash system, Dec 2008.
\newblock Accessed: 2015-07-01.

\bibitem{CasperTFG}
Ryuya Nakamura, Takayuki Jimba, and Dominik Harz.
\newblock Refinement and verification of cbc casper.
\newblock Cryptology ePrint Archive, Report 2019/415, 2019.
\newblock \url{https://eprint.iacr.org/2019/415}.

\bibitem{NSEC5}
Dimitrios Papadopoulos, Duane Wessels, Shumon Huque, Moni Naor, Jan
  V\v{c}el\'ak, Leonid Reyzin, and Sharon Goldberg.
\newblock Making {NSEC5} practical for dnssec.
\newblock IACR ePrint Report 2017/099, 2017.
\newblock \url{https://eprint.iacr.org/2017/099}.

\bibitem{sanchez2017proportional}
Luis S{\'a}nchez-Fern{\'a}ndez, Edith Elkind, Martin Lackner, Norberto
  Fern{\'a}ndez, Jes{\'u}s~A Fisteus, Pablo~Basanta Val, and Piotr Skowron.
\newblock Proportional justified representation.
\newblock In {\em Thirty-First AAAI Conference on Artificial Intelligence},
  2017.

\bibitem{sanchez2016maximin}
Luis S{\'a}nchez-Fern{\'a}ndez, Norberto Fern{\'a}ndez, Jes{\'u}s~A Fisteus,
  and Markus Brill.
\newblock The maximin support method: An extension of the d'hondt method to
  approval-based multiwinner elections.
\newblock {\em arXiv preprint arXiv:1609.05370}, 2016.

\bibitem{aura}
Elaine Shi.
\newblock Analysis of deterministic longest-chain protocols.
\newblock In {\em 2019 IEEE 32nd Computer Security Foundations Symposium
  (CSF)}, pages 122--12213. IEEE, 2019.

\bibitem{2018:Stewart:Grandpa}
Alistair Stewart.
\newblock Byzantine finality gadgets.
\newblock {\em Technical Report}, 2018.
\newblock \url{https://github.com/w3f/consensus/blob/master/pdf/grandpa.pdf}.

\bibitem{2016:Wood:Polkadot}
Gavin Wood.
\newblock Polkadot: Vision for a heterogeneous multi-chain framework.
\newblock {\em White Paper}, 2016.

\bibitem{CasperCBC}
Vlad Zamfir.
\newblock Casper the friendly ghost: A “correct-by-construction” blockchain
  consensus protocol.
\newblock 2017.

\bibitem{Zamyatin:2019:XClaim}
Alexei Zamyatin, Dominik Harz, Joshua Lind, Panayiotis Panayiotou, Arthur
  Gervais, and William~J. Knottenbelt.
\newblock {XCLAIM:} trustless, interoperable, cryptocurrency-backed assets.
\newblock In {\em 2019 {IEEE} Symposium on Security and Privacy, {SP} 2019, San
  Francisco, CA, USA, May 19-23, 2019}, pages 193--210. {IEEE}, 2019.

\end{thebibliography}
\begin{appendix}
\section{Appendix}

\subsection{SPREE} \label{sec:SPREE}

SPREE (Shared Protected Runtime Execution Enclaves) is a way for parachains to have shared code, and furthermore for the execution and state of that code to be sandboxed. From the point of view of parachain A, how much can it trust parachain B? Polkadot's shared security guarantees the correct execution of B's code with as much security as it does A's code. However, if we do not know B's code itself and even if we know the code now, maybe the governance mechanism of B can change the code and we do not trust that. This changes if we knew some of B's code, that it's governance did not have control of, and which could be sent messages by A. Then we would know how B's code would act on those messages if it was executed correctly and so shared security gives us the guarantees we need.

A SPREE module is a piece of code placed in the relay chain, that parachains can opt into. This code is part of that chains state transition validation function (STVF). The execution and state of this SPREE module are sandboxed away from the rest of the STVF's execution. SPREE modules on a remote chain can be addressed by XCMP. The distribution of messages received by a parachain would itself be controlled by a SPREE module (which would be compulsory for chains that want to use any SPREE modules).

We expect that most messages sent by XCMP will be from a SPREE module on one chain to the same SPREE module on another chain. When SPREE modules are upgraded, which involves putting updated code on the relay chain and scheduling an update block number, it is upgraded on all parachains in their next blocks. This is done in such a way as to guarantee that messages sent by a version of the SPREE module one one chain to the same module on another are never received by past versions. Thus message formats for such messages do not need to be forward compatible and we do not need standards for these formats.

For an example of the security guarantees we get from SPREE, if A has a native token, the A token, what we would like is to be sure that parachain B could not mint this token. We could enforce this by A keeping an account for B in A's state. However if an account on B want's to send some A token to a third parachain C, then it would need to inform A. A SPREE module for tokens would allow this kind of token transfer without this accounting. The module on A would just send a message to the same module on B, sending the tokens to some account. B could then send them on to C and C to A in a similar way. The module itself would account for tokens in accounts on chain B, and Polkadot's shared security and the module's code would enforce that B could never mint A tokens. XCMP's guarantee that messages will be delivered and SPREE'S guarantee that they will be interpreted correctly mean that this can be done by just sending one message per transfer and is trust free. This has applications far beyond token transfer and means that trust minimising protocols are far easier to design.

Parts of SPREEs design and implementation have yet to be fully designed. Credit goes to the reddit user u/Tawaren for the initial idea behind SPREE.

\subsection{Interoperability with External Chains}\label{sec:bridge}

Polkadot is going to host a number of bridge components to other chains. This section will be focused on bridging to BTC and ETH (1.x) and hence will mostly be reviewing bridging proof of work chains. Our bridge design is inspired by XClaim \cite{Zamyatin:2019:XClaim}.
The bridge logic will have two important parts: a bridge relay, which understands as much as possible the consensus of the bridged chain, and a bank (name change possible for PR reasons), which involves staked actors owning bridged chain tokens on behalf of Polkadot.
The bridge relay needs to be able to carry out consensus verification of the bridged chain and verify transaction inclusion proofs there. On the one hand, the bank can be used by users on the bridged chain to lock tokens as backing for the corresponding asset they want to receive on Polkadot, e.g., PolkaETH or PolkaBTC. On the other hand, users can use the bank to redeem these assets into the bridged chain tokens.
The bridge relay aims to put as much of the logic of a light/thin client of a bridged chain on a bridge parachain as is feasible – think BTC-Relay. However, crypto and storage are much cheaper on a parachain than in an ETH smart contract. We aim to put all block headers and proofs-of-inclusion of certain transactions of the bridged chain in the blocks of the bridge parachain. This is enough to decide whether a transaction is in a chain which is probably final. The idea for the bridge relay for Bitcoin and ETH1.0 is to have a longest-chain bridge chain where conflicts are resolved with a voting/attestation scheme.

\subsection{Comparison with other multi-chain systems}\label{sec:comparison}
\subsubsection{ETH2.0}
Ethereum 2.0 promises a partial transition to proof-of-stake and to deploy sharding to improve speed and throughput.  There are extensive similarities between the Polkadot and Ethereum 2.0 designs, including similar block production and finality gadgets.  

All shards in Ethereum 2.0 operate as homogeneous smart contract based chains, while parachains in Polkadot are independent heterogeneous blockchains, only some of which support different smart contract languages.  
At first blush, this simplifies deployment on Ethereum 2.0, but ``yanking'' contracts between shards dramatically complicates the Ethereum 2.0 design.  We have a smart contract language Ink! that exists so that smart contract code can more easily be migrated into being parachain code.  We assert that parachains inherent focus upon their own infrastructure should support higher performance far more easily than smart contracts.

Ethereum 2.0 asks that validators stake exactly 32 ETH, while Polkadot fixes one target number of validators, and attempts to maximise the backing stake with NPoS (see Section~\ref{sec:validators}).  At a theoretical level, we believe the 32 ETH approach results in validators being less ``independent'' than NPoS, which weakens security assumptions throughout the protocol.  We acknowledge however that Gini coefficient matters here, which gives Ethereum 2.0 an initial advantage in ``independence''.  We hope NPoS also enables more participation by Dot holders will balances below 32 ETH too.

Ethereum 2.0 has no exact analog of Polkadot's availability and validity protocol (see Section \ref{sec:validity-and-availability}).  We did however get the idea to use erasure codes from the Ethereum proposal \cite{availabilityETH2}, which aims at convincing lite clients.  

Validators in Ethereum 2.0 are assigned to each shard for attesting block of shards as parachain validators in Polkadot thus constitute the committee of the shard. The committee members receive a Merkle proof of randomly chosen code piece from a full node of the shard and verify them. If all pieces are verified and no fraud-proof is announced, then the block is considered as valid. The security of this scheme is based on having an honest majority in the committee while the security of Polkadot's scheme based on having at least one honest validator either among parachain validators or secondary checkers (see Section~\ref{sec:validity-and-availability}). Therefore, the committee size in Ethereum 2.0 is considerably large comparing to the size of parachain validators in Polkadot.

The beacon chain in Ethereum 2.0 is a proof-of-stake protocol as Polkadot's relay chain. Similarly, it has a finality gadget called Casper \cite{CasperFFG,CasperCBC} as GRANDPA in Polkadot. Casper also combines  eventual finality and  Byzantine agreement as GRANDPA but GRANDPA gives better liveness property than Casper \cite{2018:Stewart:Grandpa}.

\subsubsection{Sidechains}
An alternative way to scale blockchain technologies are using side-chains \footnote{that allow tokens from one blockchain to be considered valid on an independent blockchain and be used there}. These solutions are also addressing interoperability, in that they enabling bridging side chains to the main chain. For example, for Eth1.0 many side-chains were introduced that contributed to scalability such as Plasma Cash and Loom \footnote{https://loomx.io}.
A prominent solution that is solely based on bridging independent chains to each other is Cosmos \footnote{https://cosmos.network} that is reviewed and compared to Polkadot next.


\subsubsection{Cosmos}

Cosmos is a system designed to solve the blockchain interoperability problem that is fundamental to improve the scalability for the decentralised web. In this sense, there are surface similarities between the two systems. Hence, Cosmos consists of components which play similar roles and resemble the sub-components of Polkadot. For example, the Cosmos Hub is used to transfer messages between Comos' zones similarly to how the Polkadot Relay Chain oversees the passing of messages among Polkadot parachains.

There are however significant differences between the two systems. Most importantly, while the Polkadot system as a whole is a sharded state machine (see Section \ref{sec:relaychain}), Cosmos does not attempt to unify the state among the zones and so the state of individual zones is not reflected in the Hub's state. As the result, unlike Polkadot, Cosmos does not offer shared security among the zones. Consequently, the Cosmos cross-chain messages, are no longer trust-less. That is to say, that a receiver zone needs to fully trust the sender zone in order to act upon messages it receives from the sender. If one considers Cosmos system as a whole, including all zones in a similar way one analyses the Polkadot system, the security of such a system is equal to the security of the least secure zone. Similarly the security promise of Polkadot guarantees that validated parachain data are available at a later time for retrieval and audit (see Section \ref{sec:validity-and-availability}). In the case of Cosmos, the users are ought to trust the zone operators to keep the history of the chain state.

It is noteworthy that using the SPREE modules, Polkadot offers even stronger security than the shared security.
When a parachain signs up for a SPREE module, Polkadot guarantees that certain XCMP messages received by that parachain are being processed by the pre-defined SPREE module set of code. No similar cross-zone trust framework is offered by the Cosmos system.

Another significant difference between Cosmos and Polkadot consists in the way the blocks are produced and finalised. In Polkadot, because all parachain states are strongly connected to relay chain states, the parachain can temporarily fork alongside the relay chain. This allows the block production to decouple from the finality logic. In this sense, the Polkadot blocks can be produced over unfinalised blocks and multiple blocks can be finalised at once. On the other hand, the Cosmos zone depends on the instant finality of the Hub's state to perform a cross-chain operation and therefore a delayed finalisation halts the cross-zone operations.

\section{Glossary}

\begin{longtable}{p{.15\textwidth}p{.55\textwidth}p{.1\textwidth}p{.1\textwidth}} \label{t:time}
	\textbf{Name}  & \textbf{Description} & \textbf{Symbol} (plural)& \textbf{Def} \\
	\hline
	BABE & A mechanism to assign elected validators randomly to block production for a certain slot. && \ref{sec:babe} \\
	BABE Slot & A period for which a relay chain block can be produced. It's about 5 seconds. & \slot & \ref{sec:babe} \\
	Collator & Assist validators in block production. A set of collators is defined as \Col . & \col (\Col) & \ref{par:collators} \\
	Dot & The Polkadot native token. && \ref{sec:economics} \\
	Elected\newline- validators & A set of elected validators. & \Val & \\
	Epoch & A period for which randomness is generated by BABE. It's about 4 hours. & \ep & \\
	Era & A period for which a new validator set is decided. It's about 1 day. && \\
	Extrinsics & Input data supplied to the Relay Chain to transition states. && \ref{par:extrinsics} \\
	Fishermen & Monitors the network for misbehaviour. && \ref{par:fishermen} \\
	Gossiping & Broadcast every newly received message to peers. && \ref{sec:gossiping} \\
	GRANDPA & Mechanism to finalize blocks. && \ref{sec:grandpa} \\
	GRANDPA\newline- Round & A state of the GRANDPA algorithm which leads to block finalisation. && \ref{sec:grandpa} \\
	Nominator & Stake-holding party who nominates validators to be elected. A set of nominators is defined as \Nom . & \nom (\Nom) & \ref{par:nominators} \\
	NPoS & \emph{Nominated Proof-of-Stake} - Polkadot's version of PoS, where nominated validators get elected to be able to produce blocks. && \ref{sec:validators} \\
	Parachain & Heterogeneous independent chain. & \Par & \\
	PJR & \emph{Proportional-Justified-Representation} - Ensures that validators represent as many nominator minorities as possible. && \ref{sec:validators} \\
	PoV & \emph{Proof-of-Validity} - Mechanism where a validator can verify a block without having its full state. && \ref{sec:parachainblockproduction} \\
	Relay\newline- Chain & Ensures global consensus among parachains. && \ref{sec:relaychain} \\
	Runtime & The Wasm blob which contains the state transition functions, including other core operations required by Polkadot. && \ref{par:state_transition} \\
	Sentry\newline- nodes & Specialized proxy server which forward traffic to/from the validator. && \\
	Session & A session is a period of time that has a constant set of validators. Validators can only join or exit the validator set at a session change. && \\
	STVF & \emph{State-Transition-Validation-Function} - A function of the Runtime to verify the PoV. && \ref{sec:parachainblockproduction} \\
	Validator & The elected and highest in charge party who has a chance of being selected by BABE to produce a block. A set of candidate validators is defined as \Can . The number of validators to elect is defined as \nval . & \val (\Val)& \ref{par:validators} \\
	VRF & \emph{Verifiable-Random-Function} - Cryptographic function for determining elected validators for block production. && \ref{sec:babe} \\
	XCMP & A protocol that parachains use to send messages to each other. && \ref{sec:XCMP} \\
	\caption{Glossary for Polkadot}
\end{longtable}


\end{appendix}

\end{document}